\documentclass[10pt,showpacs,letterpaper,aps,twocolumn,nofootinbib,prx]{revtex4-1} 
\usepackage[draft]{hyperref}

\usepackage{amsthm,amsmath,amssymb,graphicx,comment} 
\usepackage{color}
\usepackage{bbold}
\usepackage{hyperref}
\usepackage[none]{hyphenat}

\newcommand{\ket}[1]{\left| #1 \right\rangle}
\newcommand{\bra}[1]{\left\langle #1 \right|}
\newcommand{\braket}[1]{\left\langle #1 \right\rangle}

\newcommand{\beq}{\begin{equation}}
\newcommand{\eeq}{\end{equation}}
\newcommand{\bea}{\begin{align}}
\newcommand{\eea}{\end{align}}

\begin{document}

\title{Contextual advantage for state discrimination}
\author{David Schmid and Robert W. Spekkens}
\affiliation{Perimeter Institute for Theoretical Physics, 31 Caroline Street North, Waterloo, Ontario Canada N2L 2Y5}
\affiliation{Institute for Quantum Computing and Department of Physics and Astronomy, University of Waterloo, Waterloo, Ontario N2L 3G1, Canada}
\email{dschmid@perimeterinstitute.ca}

\begin{abstract}
Finding quantitative aspects of quantum phenomena which cannot be explained by any classical model has foundational importance for understanding the boundary between classical and quantum theory. It also has practical significance for identifying information processing tasks for which those phenomena provide a quantum advantage. Using the framework of generalized noncontextuality as our notion of classicality, we find one such nonclassical feature within the phenomenology of quantum minimum error state discrimination. Namely, we identify quantitative limits on the success probability for minimum error state discrimination in any experiment described by a noncontextual ontological model. These constraints constitute noncontextuality inequalities that are violated by quantum theory, and this violation implies a quantum advantage for state discrimination relative to noncontextual models. Furthermore, our noncontextuality inequalities are robust to noise and are operationally formulated, so that any experimental violation of the inequalities is a witness of contextuality, independently of the validity of quantum theory. Along the way, we introduce new methods for analyzing noncontextuality scenarios, and demonstrate a tight connection between our minimum error state discrimination scenario and a Bell scenario.
\end{abstract}

\maketitle


Understanding the boundary between the quantum and the classical is of fundamental importance for understanding quantum theory. One successful metric for nonclassicality, violation of Bell's notion of local causality~\cite{Bell}, defines a clear departure from classicality in relativistic theories, but is relevant only for experiments with space-like separated measurements. Another notion of classicality, which concerns context-independence,
 was proposed by Kochen-Specker~\cite{KS} and Bell~\cite{Bell2}, and has since been significantly refined and generalized~\cite{gencontext}. It is the generalized notion of noncontextuality from Ref.~\cite{gencontext} which we study in this paper, but we refer to it simply as ``noncontextuality'' hereafter. As a metric for nonclassicality, the failure of noncontextuality has a broader scope than the failure of local causality insofar as it does not require space-like separation. It has also been shown to subsume many other pre-existing notions of nonclassicality, such as the negativity of quasi-probability representations~\cite{negativity}, the generation of anomalous weak values~\cite{AWV}, and even the aforementioned violations of local causality~\cite{gencontext}. 

The quantum-classical boundary is also of practical importance in identifying tasks which admit of a quantum advantage. For example, violations of Bell inequalities have been shown to be a resource for device-independent key distribution~\cite{DIQKD}, certified randomness~\cite{randomness}, and communication complexity~\cite{comm}. 
The failure of noncontextuality has also been shown to be a resource, leading to advantages for cryptography~\cite{POM,RAC,RAC2} and computation~\cite{magic,comp1,comp2}.

We here analyze minimum-error state discrimination (MESD) from the point of view of noncontextuality.
Quantum state discrimination is a task wherein one must guess which quantum state describes a given quantum system when the state of that system is drawn from a known set of possibilities with a known prior distribution, and the estimation is based on the outcome of a measurement of one's choosing.  In the ``minimum error'' variety of state discrimination,  the objective is to minimize the probability that the estimate is in error.   We here focus on the simplest case of a set containing just two states having equal a priori probability. 


Although it is common to assert that the impossibility of perfectly discriminating nonorthogonal quantum states is an intrinsically nonclassical effect, this claim does not meet the minimal standard that one should require of {\em any} claim that some operational feature of quantum theory cannot be explained classically: namely, that it be justified by a rigorous no-go theorem. Such a theorem articulates a principle of classicality which has implications for operational statistics, and then proves that these implications are inconsistent with some operational feature(s) of quantum theory. Because the principle of noncontextuality constrains operational statistics and also has very broad scope, it is a particularly useful notion of classicality. 
If one does take it 
as one's principle of classicality, then the impossibility of discriminating nonorthogonal pure quantum states \textit{cannot} be considered a nonclassical effect because there are subtheories of quantum theory (containing a strict subset of the states, measurements and transformations of the full theory)~\cite{epistricted} wherein this phenomenon arises and which admit of a noncontextual model.  (Within such models, the phenomenon can be attributed to the fact that the probability distributions associated to such quantum states are overlapping\footnote{Such models are 
 $\psi$-epistemic, in the terminology of Ref.~\cite{Harrigan}.}.)  
It follows that one must look at more nuanced aspects of the phenomenology of quantum state discrimination to identify features which are truly nonclassical by these lights.

We identify one such strongly nonclassical aspect of minimum error state discrimination: the particular dependence of the probability of successful discrimination on the overlap of the quantum states. For a given overlap, the quantum probability of discrimination is larger than can be accounted for by a noncontextual model. After presenting this result as a no-go theorem---that no noncontextual model can reproduce certain features of quantum MESD---we reformulate the problem in a manner which makes no reference to quantum theory, and which does not rely on any theoretical idealizations such as noise-free measurements or preparations.  
Our entirely operational formulation allows us to derive inequalities which can experimentally witness a contextual advantage for state discrimination, in the presence of noise and independently of the validity of quantum theory. 

Our result identifies a key feature of quantum state discrimination which cannot be understood in any noncontextual model, and hence which is strongly nonclassical. Because quantum state discrimination is a primitive in many important quantum information processing protocols~\cite{Bennett,zeroerror}, this work constitutes a first step towards identifying contextuality as a resource for more tasks concerning communication, computation, and cryptography. 

We also prove an isomorphism between our operational MESD scenario and a two-party Bell test in which one party performs one of a pair of binary-outcome measurements and the other performs one of three binary-outcome measurements. 
This is similar to the fact that the noncontextuality inequality delimiting the success rate for parity-oblivious multiplexing~\cite{POM} is isomorphic to the CHSH inequality in the Bell scenario~\cite{POM}.

Finally, we introduce two powerful new technical tools. First, we generalize existing methods for simulating exact operational equivalences~\cite{unwarranted}. Namely, while Ref.~\cite{unwarranted} shows how one may find a set of procedures which respects certain operational equivalences exactly, we have further demonstrated that one can find procedures which respect operational equivalences and \textit{simultaneously} obey useful auxiliary constraints, such as the symmetries native to our ideal MESD scenario. This tool may have more general applications in the comparison of experimental data with theoretical expectations. More importantly, we find our noncontextuality inequalities using a novel algorithm (presented in Appendix~\ref{NCnoise}) for deriving the full set of necessary and sufficient noncontextuality inequalities for \textit{any} finite prepare-and-measure scenario, with respect to any fixed operational equivalences\footnote{A full description of this algorithm can be found in Ref.~\cite{all}.}.


\section{Operational Theories and Ontological Models}

An operational theory is a specification of sets of primitive laboratory operations (e.g., preparations and measurements) and a prescription for finding the probabilities $p(k|M,P)$ for each outcome $k$ given any measurement $M$ performed on any preparation $P$. Two preparations $P$ and $P'$ are termed \textit{operationally equivalent} if they cannot be differentiated by the statistics of any measurement; we denote this operational equivalence by 
\beq
P \simeq P'.
\eeq  
In this article, quantum theory is understood as an operational theory.  In the quantum formalism, the density operator specifies the statistics for all measurements, so that two preparation procedures are operationally equivalent if and only if they are represented by the same density operator. 

An ontological model of an operational theory has the following form. To every system, there is associated an ontic state space $\Lambda$, where each ontic state $\lambda \in \Lambda$ specifies all the physical properties of the system.  Each preparation $P$ of a system is presumed to sample the system's ontic state $\lambda$ at random from a probability distribution, denoted $\mu_P(\lambda)$ and termed the \textit{epistemic state} associated to $P$, where
\begin{align}
\forall \lambda: \quad 0 \leq \mu_P(\lambda), \label{prob1} \\
\int_{\Lambda} d\lambda \, \mu_P(\lambda) = 1 \label{norm1}.
\end{align}
Each measurement $M$ on a system is presumed to have its outcome $k$ sampled at random in a manner that depends on the ontic state $\lambda$. The term {\em effect} will be used to refer to the pair consisting of a measurement, $M$, together with one of its outcomes, $k$, and will be denoted by $k|M$.  The probability of outcome $k$ given measurement $M$, considered as a function of $\lambda$, will be termed the {\em response function} associated to $k|M$, and denoted $\xi_{k|M}(\lambda)$, where
\begin{align}
\forall \lambda, \forall k: \quad 0 \leq \xi_{k|M}(\lambda), \\
\forall \lambda: \quad \sum_k \xi_{k|M}(\lambda) = 1 .
\end{align}
Finally, an ontological model of an operational theory must reproduce the latter's empirical predictions; that is,
\begin{equation} \label{probs}
p(k | M,P) = \int_{\Lambda} d\lambda \, \xi_{k|M}(\lambda) \mu_P(\lambda).
\end{equation}

We are now in a position to describe the assumption of preparation noncontextuality defined in Ref.~\cite{gencontext}.
An ontological model is said to be \textit{preparation noncontextual} if it assigns the same epistemic state to all operationally equivalent preparations~\cite{gencontext}:
\beq \label{PNCdefn}
P \simeq P' \implies \mu_P(\lambda) = \mu_{P'}(\lambda).
\eeq
In operational quantum theory, the principle of preparation noncontextuality is respected whenever any two preparations that are associated to the same density operator are represented by the same epistemic state. For instance, different ensembles of states that average to the same mixed state (and for which one discards the information about which element of the ensemble was prepared) are operationally equivalent, and must be assigned the same epistemic state in a preparation noncontextual model.  

Although there is a corresponding notion of measurement noncontextuality (namely, that operationally equivalent outcomes of measurements are represented by the same response functions), we will not have use of it in this article.

A few terminological conventions will be useful.
A measurement is said to be represented as \textit{outcome-deterministic} in the ontological model if the associated response functions all take values in $\{0,1\}$. The support of an epistemic state is defined as the set of $\lambda \in \Lambda$ which are assigned nonzero probability by it, $\text{supp}[\mu_P(\lambda)] \equiv \{ \lambda : \mu_P(\lambda) \neq 0  \}$, while the support of a response function is defined as the set of $\lambda \in \Lambda$ for which the response function is nonzero, $\text{supp}[\xi_{k|M}(\lambda)] \equiv \{ \lambda: \xi_{k|M}(\lambda) \neq 0  \}$.

\section{Quantum Minimum Error State Discrimination} \label{QMESD}

\begin{figure}[h] 
\centering
\includegraphics[width=0.5\textwidth]{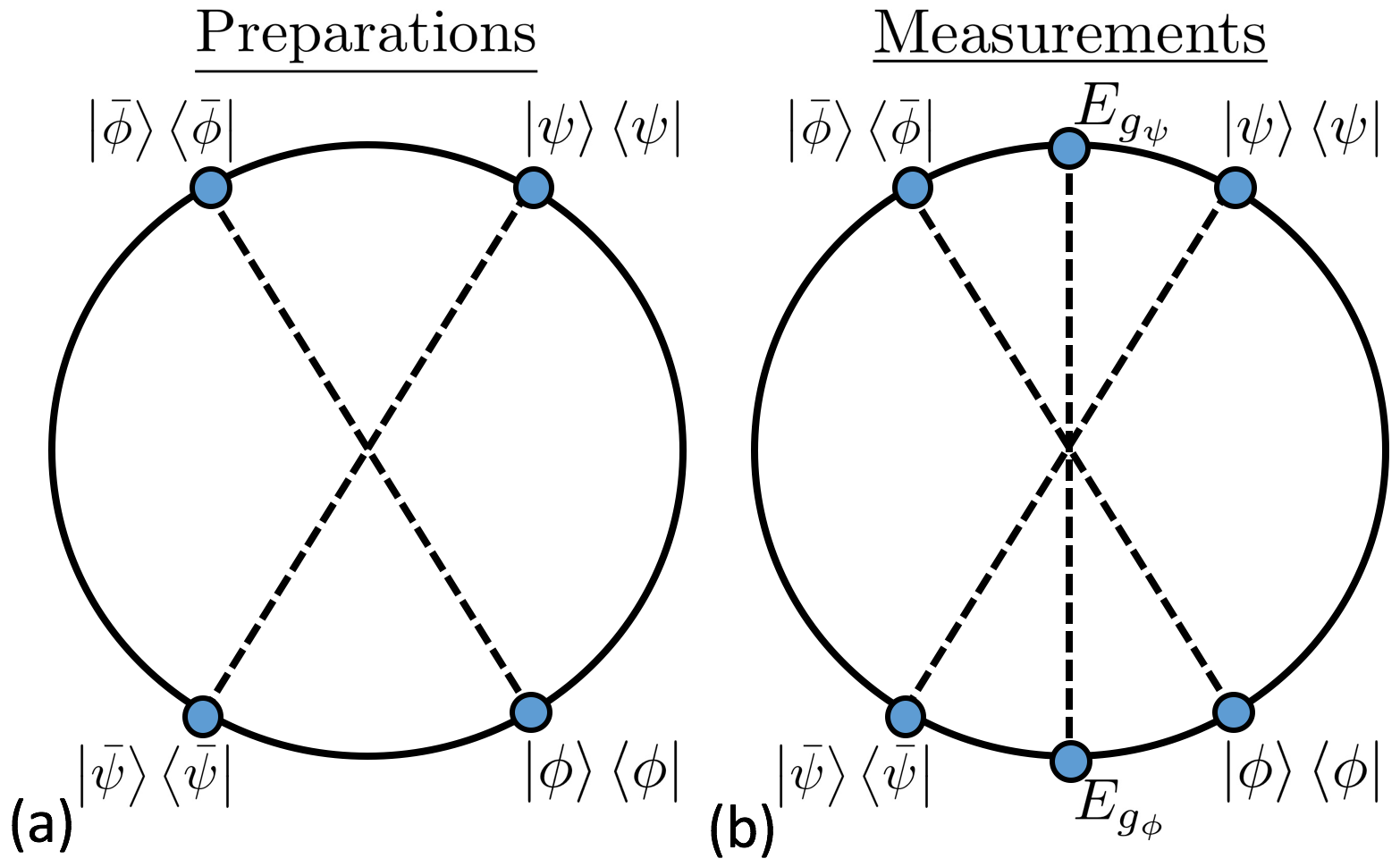}
\caption{The quantum states and measurements in our scenario, depicted as Bloch vectors in an equatorial plane of the Bloch ball.} \label{ball} 
\end{figure}

We begin with the problem of discriminating two nonorthogonal pure quantum states $\ket{\phi}$ and $\ket{\psi}$. These two states span a 2-dimensional space, so we can represent them as points in an equatorial plane of the Bloch ball, as in Fig.~\ref{ball}. 

First, we consider the operational signature of their nonorthogonality. A measurement of the $\phi$ basis, $B_{\phi} \equiv \{ \ket{\phi}\bra{\phi}, \ket{\bar{\phi}}\bra{\bar{\phi}} \},$ perfectly distinguishes between state $\ket{\phi}$ and its complement; we denote the associated outcomes by $\phi$ and $\bar{\phi}$, respectively. A measurement of the $\psi$ basis, $ B_{\psi} \equiv \{\ket{\psi}\bra{\psi}, \ket{\bar{\psi}}\bra{\bar{\psi}} \}$, does the same for the state $\ket{\psi}$ and its complement, with associated outcomes $\psi$ and $\bar{\psi}$.  
If one implements the $\psi$ basis measurement on the state $\phi$, the probability of obtaining the $\psi$ outcome is
\beq \label{defc}
c_q = \mathrm{Tr}[\ket{\phi} \braket{\phi|\psi} \bra{\psi}] = |\braket{\phi|\psi}|^2,
\eeq
Because one could think of this quantity as the probability that $\phi$ passes the test for $\psi$ and thus is confusable with $\psi$,  we henceforth call it the {\em confusability}.  
Note that if one implements the $\phi$ basis measurement on the state $\psi$, the probability of obtaining the $\phi$ outcome is also $c_q$. 

If $\ket{\phi}$ and $\ket{\psi}$ have nonzero confusability (i.e., if they are not orthogonal), then no measurement can distinguish between the two without incurring a nonzero probability of error. We denote the discriminating measurement by 
$B_d \equiv \{ E_{g_{\phi}}, E_{g_{\psi}} \},$ where the outcome for which one should guess $\phi$
 (respectively $\psi$)
 is denoted $g_{\phi}$ (respectively $g_{\psi}$).
Assuming equal prior probabilities of $\ket{\phi}$ and $\ket{\psi}$, the probability of guessing the state correctly with this measurement is 
\beq
s_q \equiv \frac{1}{2}\mathrm{Tr}[E_{g_{\phi}} \ket{\phi} \bra{\phi}] + \frac{1}{2}\mathrm{Tr}[E_{g_{\psi}} \ket{\psi} \bra{\psi}].
\eeq
We assume that the discriminating measurement has the natural symmetry property $\mathrm{Tr}[E_{g_{\phi}} \ket{\phi} \bra{\phi}] = \mathrm{Tr}[E_{g_{\psi}} \ket{\psi} \bra{\psi}]$ so that 
\beq \label{defs}
s_q =\mathrm{Tr}[E_{g_{\phi}} \ket{\phi} \bra{\phi}] = \mathrm{Tr}[E_{g_{\psi}} \ket{\psi} \bra{\psi}].
\eeq


The measurement scheme that yields the greatest probability of guessing correctly which of two nonorthogonal states was prepared is called the \textit{minimum error} state discrimination (MESD) scheme. 
Since $\ket{\phi}$ and $\ket{\psi}$ are prepared with equal probability, the POVM $\{ E_{g_{\phi}}, E_{g_{\psi}} \}$ achieving MESD is the one consisting of projectors onto the basis that straddles $\ket{\phi}$ and $\ket{\psi}$ in Hilbert space, which is depicted in the Bloch sphere in Fig.~\ref{ball}.  This is called the {\em Helstrom measurement}~\cite{Helstrom}. It is well-known that the probability of guessing the state correctly using the Helstrom measurement is
\beq \label{sdef}
s_q = \frac{1}{2}(1+\sqrt{1-|\braket{\phi|\psi}|^2}) = \frac{1}{2}(1+\sqrt{1-c_q}).
\eeq 

We have now described all of the preparations and measurements that usually appear in a discussion of the problem of discriminating two nonorthogonal quantum states, and some basic facts about the relations that hold among the operational quantities characterizing the discrimination problem (i.e., facts about the phenomenology of quantum state discrimination).  However, these facts are insufficient for deriving a no-go theorem for noncontextuality. The reason is that the preparations and measurements described thus far do not exhibit any operational equivalences via which the assumption of noncontextuality could imply nontrivial constraints on the ontological model.  


However, there is a simple solution: we also consider the problem of discriminating the pair of quantum states that are complementary to $\ket{\phi}$ and $\ket{\psi}$, namely, $\ket{\bar{\phi}}$ and $\ket{\bar{\psi}}$, also depicted in Fig.~\ref{ball}. By symmetry, the confusability of $\ket{\bar{\phi}}$ and $\ket{\bar{\psi}}$ is also equal to $c_q$, and the success rate for distinguishing $\ket{\bar{\phi}}$ and $\ket{\bar{\psi}}$ when they have equal prior probability is also equal to $s_q$ (where the optimal measurement is again $\{ E_{g_{\phi}}, E_{g_{\psi}} \},$ but now the outcomes $g_\phi$ and $g_\psi$ signal one to guess preparations $\ket{\bar{\psi}}$ and $\ket{\bar{\phi}}$, respectively).  So the $\ket{\bar{\phi}}$ vs. $\ket{\bar{\psi}}$ discrimination problem is a mirror image of the $\ket{\phi}$ vs. $\ket{\psi}$ discrimination problem, and consequently does not require specifying any additional facts about the phenomology of quantum state discrimination.  However, the inclusion of $\ket{\bar{\phi}}$ and $\ket{\bar{\psi}}$ in our analysis provides us with a nontrivial operational equivalence relation among the preparations, namely, 
\beq \label{mix}
\frac{1}{2} \ket{\phi}\bra{\phi} + \frac{1}{2} \ket{\bar{\phi}}\bra{\bar{\phi}} = \frac{1}{2} \ket{\psi}\bra{\psi} + \frac{1}{2} \ket{\bar{\psi}}\bra{\bar{\psi}} = \frac{\mathbb{1}}{2}.
\eeq
We will show that this equivalence relation together with the phenomenology of quantum state discrimination described above is sufficient to derive a no-go theorem for noncontextuality. 

The probability of a given measurement outcome occurring on a given preparation, for every possible pairing thereof, is summarized in 
Table~\ref{QMtable}.  Here, the columns correspond to the distinct state-preparations and the rows correspond to the distinct effects (where one need only include a single effect for each binary-outcome measurement given that the probability for the other effect is fixed by normalization).
\begin{center}
\begin{table} [!htb]
\newcommand*{\TitleParbox}[1]{\parbox[c]{1.75cm}{\raggedright #1}}%
 \begin{tabular}{| c | c | c | c | c |} 
 \hline
  & \parbox[c]{1.2cm}{$\ket{\phi}$} &  \parbox[c]{1.2cm}{$\ket{\psi}$} &  \parbox[c]{1.2cm}{$\ket{\bar{\phi}}$}  &  \parbox[c]{1.2cm}{$\ket{\bar{\psi}}$} \\ [0.5ex] 
 \hline 
$\ket{\phi}\bra{\phi}$ & 1 & $c_q$ & $0$ & $1-c_q$ \\ [0.2ex] 
 \hline
$\ket{\psi}\bra{\psi}$ & $c_q$ & $1$ & $1-c_q$ & 0 \\ [0.2ex] 
 \hline
 $E_{g_{\phi}}$ & $s_q$ & $1-s_q$ & $1-s_q$ & $s_q$ \\ [0.2ex] 
\hline
\end{tabular}
\caption{Data table in the ideal quantum case. 
} \label{QMtable}
\end{table}
\end{center}

\section{Noncontextuality no-go theorem for MESD in Quantum Theory} \label{ncnogo}

The fact that the ontological model must reproduce the probabilities in Table~\ref{QMtable} via Eq.~\eqref{probs} implies constraints on the epistemic states associated to the four preparations and the response functions associated to the three effects.  For instance, to reproduce the first column of the table, one requires that
\begin{align}
\int_{\Lambda} d\lambda \, \xi_{\phi|B_{\phi}}(\lambda) \mu_{\phi}(\lambda) &= 1,\label{c1}\\
\int_{\Lambda} d\lambda \, \xi_{\psi|B_{\psi}}(\lambda) \mu_{\phi}(\lambda) &= c_q,\label{c2}\\
\int_{\Lambda} d\lambda \, \xi_{g_{\phi}|B_d}(\lambda) \mu_{\phi}(\lambda) &= s_q.\label{c3}
\end{align}

Given that convex mixtures of preparations are represented in an ontological model by the corresponding mixture of epistemic states (see Eq.~(7) of~\cite{negativity} and the surrounding discussion), it follows that $\frac{1}{2} \ket{\phi}\bra{\phi} + \frac{1}{2} \ket{\bar{\phi}}\bra{\bar{\phi}}$ is represented by $\frac{1}{2}\mu_{\phi}(\lambda) + \frac{1}{2}\mu_{\bar{\phi}}(\lambda)$, and $\frac{1}{2} \ket{\psi}\bra{\psi} + \frac{1}{2} \ket{\bar{\psi}}\bra{\bar{\psi}}$ is represented by $\frac{1}{2}\mu_{\psi}(\lambda) + \frac{1}{2} \mu_{\bar{\psi}}(\lambda)$.  But because both of these mixtures of preparations are associated to the completely mixed state (Eq.~\eqref{mix}), they are operationally equivalent, and thus by the assumption of preparation noncontextuality, they are represented by the same epistemic state.  It follows that
\begin{equation} \label{PNC1}
\frac{1}{2}\mu_{\phi}(\lambda) + \frac{1}{2}\mu_{\bar{\phi}}(\lambda) = \frac{1}{2}\mu_{\psi}(\lambda) + \frac{1}{2} \mu_{\bar{\psi}}(\lambda). 
\end{equation}

Any ontological model satisfying noncontextuality, and consequently Eq.~\eqref{PNC1}, and reproducing the form of the data in Table~\ref{QMtable}, and consequently Eqs.~\eqref{c1}-\eqref{c3} and their kin, can be shown to satisfy the following trade-off between $s_q$ and $c_q$:
\beq \label{eq}
s_q \leq 1-\frac{c_q}{2}.
\eeq
An intuitive proof is provided in Section~\ref{intuitive}, where we also discuss how this result is related to the results of Refs.~\cite{Maroney,Leifer,Barrett2}. (In Appendix~\ref{sec:noiseless}, we provide a proof using more general methods, which generalizes more easily to the noisy case discussed later, in Section~\ref{sec:ineq}.)

This tradeoff relation contradicts the one known to be optimal in quantum theory, Eq.~\eqref{sdef}. The optimal quantum tradeoff generally allows \textit{higher} success rates for a given confusability than the noncontextual tradeoff. Therefore, we conclude that the phenomenology of minimum-error state discrimination in the noiseless quantum case is inconsistent with the principle of noncontextuality.

\begin{figure}[h]
\centering
\includegraphics[width=0.4\textwidth]{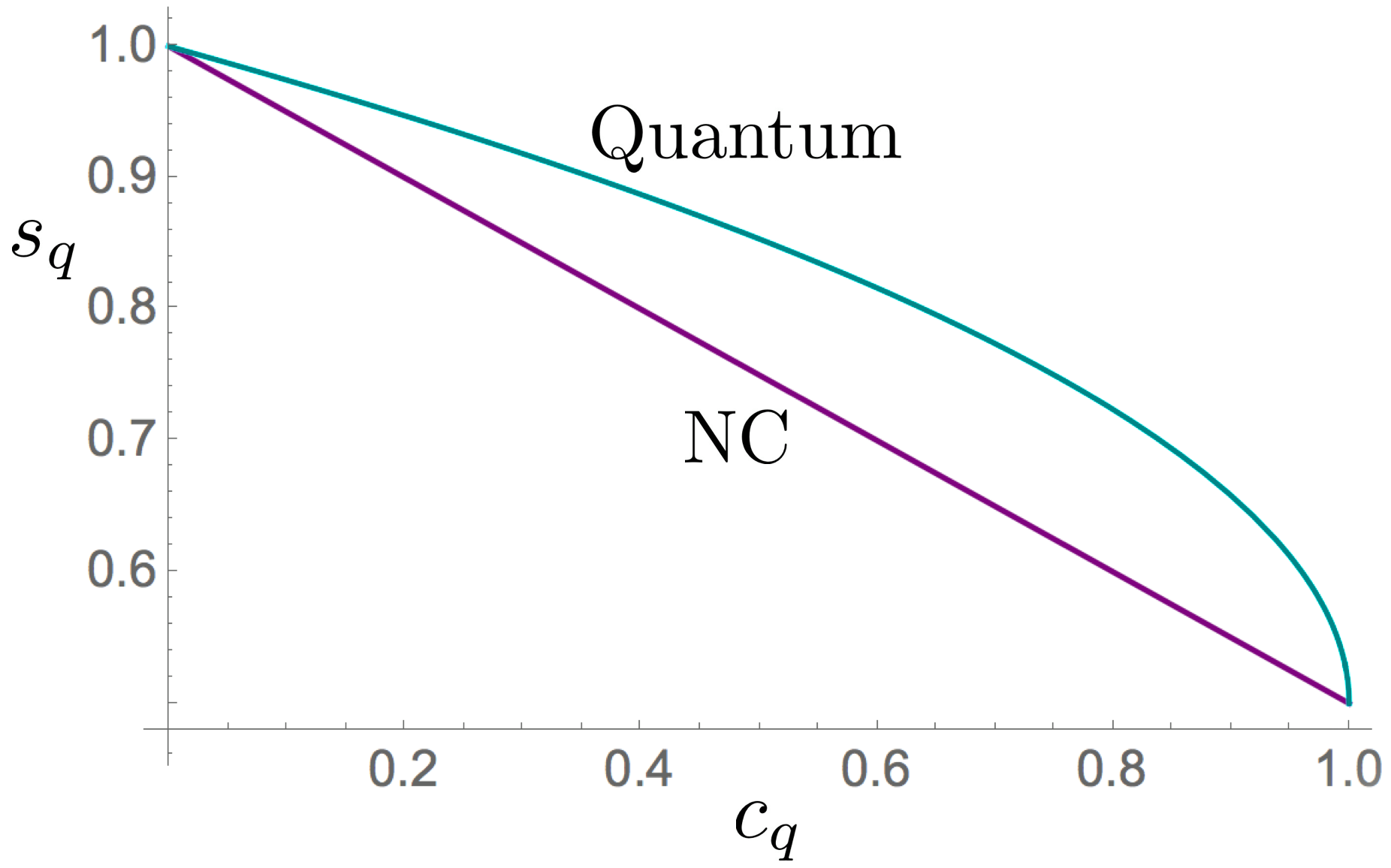}
\caption{Optimal tradeoff for a noncontextual model (purple line) and for quantum theory (light blue curve).} \label{QMandNC1}
\end{figure}

In Fig.~\ref{QMandNC1}, we plot the maximum success rate for MESD as a function of the confusability for both quantum theory (Eq.~\eqref{sdef}) and for a noncontextual model (the tradeoff that saturates the inequality of Eq.~\eqref{eq}). 

\subsection{Intuitive proof of the noncontextual tradeoff} \label{intuitive}
We now introduce some basic facts from classical probability theory, which we then leverage to prove Eq.~\eqref{eq}.

Suppose that a classical variable $\lambda$ has been sampled from one of two overlapping probability distributions, $p(\lambda|a)$ and $p(\lambda|b)$. Absent additional information, it is straightforward to see that in trying to guess which of the two distributions a given $\lambda$ was drawn from, one cannot do better than guessing `distribution $a$' for the values of $\lambda$ for which $p(a|\lambda) > p(b|\lambda)$, and guessing `distribution $b$' when the opposite is true. (Of course, it is irrelevant what one guesses for the values of $\lambda$ for which $p(a|\lambda) = p(b|\lambda)$.) In the special case we are considering, with equal prior probability $p(a)=p(b)=\frac{1}{2}$ for the two options, if we perform a Bayesian inversion, we find $p(\lambda|a) > p(\lambda|b)$ if and only if $p(a|\lambda) > p(b|\lambda)$, and hence one should guess `distribution $a$' for the values of $\lambda$ for which $p(\lambda|a) > p(\lambda|b)$, and guess `distribution $b$' when the opposite is true.


The probability that the guess $g \in \{ a, b\}$ was correct given a particular value of $\lambda$ is simply $p(g|\lambda)$. Since we always guess the distribution $a$ or $b$ that has the higher likelihood of being correct, the probability that we are right in each run is simply $\text{max}\{ p(a|\lambda),p(b|\lambda) \}$. On average, then, the success probability $r$ is 
\begin{align}
r&=\int_{\Lambda} d\lambda \  p(\lambda) \text{max}\{ p(a|\lambda),p(b|\lambda) \}\\
&= \int_{\Lambda} d\lambda \  p(\lambda) (1-\text{min}\{ p(a|\lambda),p(b|\lambda)) \} \label{condnorm}\\ 
&= 1- \int_{\Lambda} d\lambda \  \text{min}\{ p(a|\lambda)p(\lambda),p(b|\lambda)p(\lambda) \} \\
&= 1- \int_{\Lambda} d\lambda \  \text{min}\{ p(\lambda|a)p(a),p(\lambda|b)p(b) \} \\
&= 1- \frac{1}{2}\int_{\Lambda} d\lambda \ \text{min}\{ p(\lambda|a),p(\lambda|b) \},
\end{align}
where the equality on line~\eqref{condnorm} uses the fact that $p(a|\lambda)+p(b|\lambda) = 1$ for all $\lambda$.
The quantity $\int_{\Lambda} d\lambda \  \text{min} \{ p(\lambda|a), p(\lambda|b) \}$ is termed the {\em classical overlap} of the probability distributions $p(\lambda|a)$ and $p(\lambda|b)$. 

In an MESD scenario, the task is to guess, in each particular run of the experiment, whether a system was prepared by state-preparation $\ket{\phi}$ or by state-preparation $\ket{\psi}$.  If the experiment is described by an ontological model, then this task corresponds to guessing, from a single sample of the ontic state $\lambda$ of the system, whether it was sampled from the distribution $\mu_{\phi}(\lambda)$ or from $\mu_{\psi}(\lambda)$.  Given that we do not assume any operational equivalence relations among the measurements in the experiment, the assumption of measurement noncontextuality does not place any constraints on the ontological representation of the measurements.  Therefore, in particular, the Helstrom measurement is \textit{at best} represented in the ontological model by the set of response functions that yield the maximum probability of guessing which distribution the ontic state $\lambda$ was sampled from.  From our discussion concerning two overlapping classical probability distributions, it is clear that this corresponds to a measurement that returns the $g_{\phi}$ outcome whenever $\mu_{\phi}(\lambda) > \mu_{\psi}(\lambda)$ and the $g_{\psi}$ outcome whenever $\mu_{\phi}(\lambda) < \mu_{\psi}(\lambda)$, and that the probability of guessing correctly based on the outcome of the Helstrom measurement is upper bounded as follows:\footnote{A mathematically equivalent version of this upper bound was previously proven under different assumptions in Refs.~\cite{Barrett2,PBRexp}. The former article considered the assumption that this inequality is saturated as a constraint on ontological models, which they termed ``maximal $\psi$-episemicity''. (Note that this constraint is different from the constraint considered in Ref.~\cite{Leifer} even though it has the same name.)}
\beq\label{sqint}
s_q \le 1-\frac{1}{2} \int_{\Lambda} d\lambda \  \text{min} \{ \mu_{\phi}(\lambda), \mu_{\psi}(\lambda) \}.
\eeq

We will now show that in a noncontextual model, 
\beq\label{cqint}
c_q=\int_{\Lambda} d\lambda \  \text{min} \{ \mu_{\phi}(\lambda), \mu_{\psi}(\lambda) \},
\eeq
so that substituting Eq.~\eqref{cqint} into Eq.~\eqref{sqint}, we infer that $s_q \leq 1-\frac{c_q}{2}$, the noncontextual bound on the trade-off between $s_q$ and $c_q$ described in Eq.~\eqref{eq}.

Firstly, in any preparation noncontextual model the response function $\xi_{i}(\lambda)$ for a projector onto pure state $\ket{i}$ satisfies
\begin{equation} \label{ODSM}
  \xi_{i}(\lambda) = 
      \begin{cases}
        1, & \text{if } \lambda \in \text{supp}[\mu_i(\lambda)] \\
	0, & \text{otherwise}.
      \end{cases}
\end{equation}

This \textit{outcome determinism for sharp measurements} was first proven in Ref.~\cite{gencontext}. It can be seen by considering the projector as part of some projective measurement $M$ with effects 
$\{  E_i = \ket{i} \bra{i} \}$, and the corresponding basis of pure states $\{ \rho_i = \ket{i} \bra{i}\}$, 
so that $\mathrm{Tr}[E_i \rho_j ] = \delta_{i,j}$. Denoting the epistemic state of $\rho_j$ as $\mu_j(\lambda)$ and the response function for $E_i$ as $\xi_{i|M}(\lambda)$, this implies that $\int \mu_j(\lambda) \xi_{i|M}(\lambda) d\lambda = \delta_{i,j}$. Because $\mu_j(\lambda)$ is a normalized probability distribution, this implies that, for \textit{any} ontological model,
\begin{equation} \label{od}
  \xi_{i|M}(\lambda) = 
      \begin{cases}
        1, & \text{if } \lambda \in \text{supp}[\mu_i(\lambda)] \\
	0, & \text{if } \lambda \in \text{supp}[\mu_{j \neq i}(\lambda)].
      \end{cases}
\end{equation}
Eq.~\eqref{od} is not equivalent to Eq.~\eqref{ODSM}, since there may exist ontic states that are not in the support of \textit{any} of the $\mu_i(\lambda)$, and Eq.~\eqref{od} does not constrain such ontic states in any way.
In a preparation noncontextual model, however, we can furthermore show that there are no ontic states outside of the union of the supports of the set of basis states, $\cup_i \text{supp}[\mu_i(\lambda)]$, as follows. Every density operator $\rho$ appears in \textit{some} decomposition of the maximally mixed state $\frac{1}{d}\mathbb{1}$. By preparation noncontextuality, every such decomposition has the \textit{same} distribution $\mu_{\frac{1}{d}\mathbb{1}}(\lambda)$ over ontic states. Thus, every ontic state in the support of the corresponding $\mu_{\rho}(\lambda)$ \textit{also} appears in the support of $\mu_{\frac{1}{d}\mathbb{1}}(\lambda)$ , so the full state space $\Lambda$ is equivalent to $\text{supp}[\mu_{\frac{1}{d}\mathbb{1}}(\lambda)]$. Furthermore, for the basis of states $\{ \rho_i \}$ above, $\frac{1}{d} \sum_i \rho_i = \frac{1}{d}\mathbb{1}$, so preparation noncontextuality implies that $\sum_i \frac{1}{d} \mu_i (\lambda) = \mu_{\frac{1}{d}\mathbb{1}}(\lambda)$, and therefore $\cup_i \text{supp}[\mu_i(\lambda)] = \text{supp}[\mu_{\frac{1}{d}\mathbb{1}}(\lambda)] = \Lambda$.
Thus every ontic state $\lambda$ must be in the support of exactly one of the $\rho_i$, and Eq.~\eqref{od} can be strengthened to Eq.~\eqref{ODSM}.

Recalling the expression for the confusability of quantum states $\ket{\phi}$ and $\ket{\psi}$ in an ontological model, $c_q = \int_{\Lambda} d\lambda \  \xi_{\phi|B_{\phi}}(\lambda) \mu_{\psi}(\lambda)$, Eq.~\eqref{ODSM} implies that for a preparation noncontextual model:
\begin{align}
c_q &= \int_{\text{supp}[\mu_{\phi}(\lambda)]} d\lambda \  \mu_{\psi}(\lambda).\label{expforcq}
\end{align}

By virtue of the symmetry of the problem, the analogous expression with the roles of $\phi$ and $\psi$ reversed also holds. The fact that the expression for the ideal confusability $c_q = |\langle \phi | \psi \rangle|^2$ of $\phi$ and $\psi$ in a preparation-noncontextual model is given by Eq.~\eqref{expforcq} was noted by Leifer and Maroney~\cite{Leifer}.

The second implication of preparation noncontextuality which we require to prove Eq.~\eqref{cqint} is that for each of the four quantum states $\Psi \in \{\phi,\psi,\bar{\phi},\bar{\psi}\}$, 
$\mu_{\Psi}(\lambda) = 2\mu_{ \frac{\mathbb{1}}{2}}(\lambda)$ for all $\lambda \in \text{supp}[\mu_{\Psi}(\lambda)]$, where 
$\mu_{ \frac{\mathbb{1}}{2}}(\lambda)$ is the distribution associated with the maximally mixed state $\frac{\mathbb{1}}{2}$.
This was also first proven in Ref.~\cite{gencontext}, and follows immediately from preparation noncontextuality, $\frac{1}{2}\mu_{\phi}(\lambda)+ \frac{1}{2}\mu_{\bar{\phi}}(\lambda) = \frac{1}{2}\mu_{\phi}(\lambda)+\frac{1}{2}\mu_{\bar{\phi}}(\lambda) = \mu_{ \frac{\mathbb{1}}{2}}(\lambda)$, and the fact that an ontic state can be in the support of at most one state from a set of orthogonal states; that is,
$\mu_{\phi}(\lambda)\mu_{\bar{\phi}}(\lambda) = 0$ and $\mu_{\psi}(\lambda)\mu_{\bar{\psi}}(\lambda) = 0$.

Hence for all $\lambda \in \text{supp}[\mu_{\phi}(\lambda)] \cap \text{supp}[\mu_{\psi}(\lambda)]$, we have $\mu_{\phi}(\lambda)=\mu_{\psi}(\lambda) = 2\mu_{ \frac{\mathbb{1}}{2}}(\lambda)$.
It follows that $\text{min} \{ \mu_{\phi}(\lambda), \mu_{\psi}(\lambda) \} = \mu_{\phi}(\lambda)=\mu_{\psi}(\lambda)$ for all $\lambda \in \text{supp}[\mu_{\phi}(\lambda)] \cap \text{supp}[\mu_{\psi}(\lambda)]$, and is equal to $0$ everywhere else, and consequently
\begin{align} \label{preepist}
&\int_{\text{supp}[\mu_{\phi}(\lambda)]} d\lambda \ \mu_{\psi}(\lambda) \nonumber \\
&\quad =\int_{\text{supp}[\mu_{\phi}(\lambda)] \cap \text{supp}[\mu_{\psi}(\lambda)] } d\lambda \ \mu_{\psi}(\lambda) \nonumber \\
&\quad =\int_{\text{supp}[\mu_{\phi}(\lambda)] \cap \text{supp}[\mu_{\psi}(\lambda)]} d\lambda \  \text{min} \{ \mu_{\phi}(\lambda), \mu_{\psi}(\lambda) \} \nonumber  \\
&\quad =\int_{\Lambda} d\lambda \  \text{min} \{ \mu_{\phi}(\lambda), \mu_{\psi}(\lambda) \}, 
\end{align}
Finally, Eq.~\eqref{expforcq} and Eq.~\eqref{preepist} together imply Eq.~\eqref{cqint}, which is what we sought to prove.

\subsection{Graphical summary of the proof}

\begin{figure}[!htb] 
\centering
\includegraphics[width=0.5\textwidth]{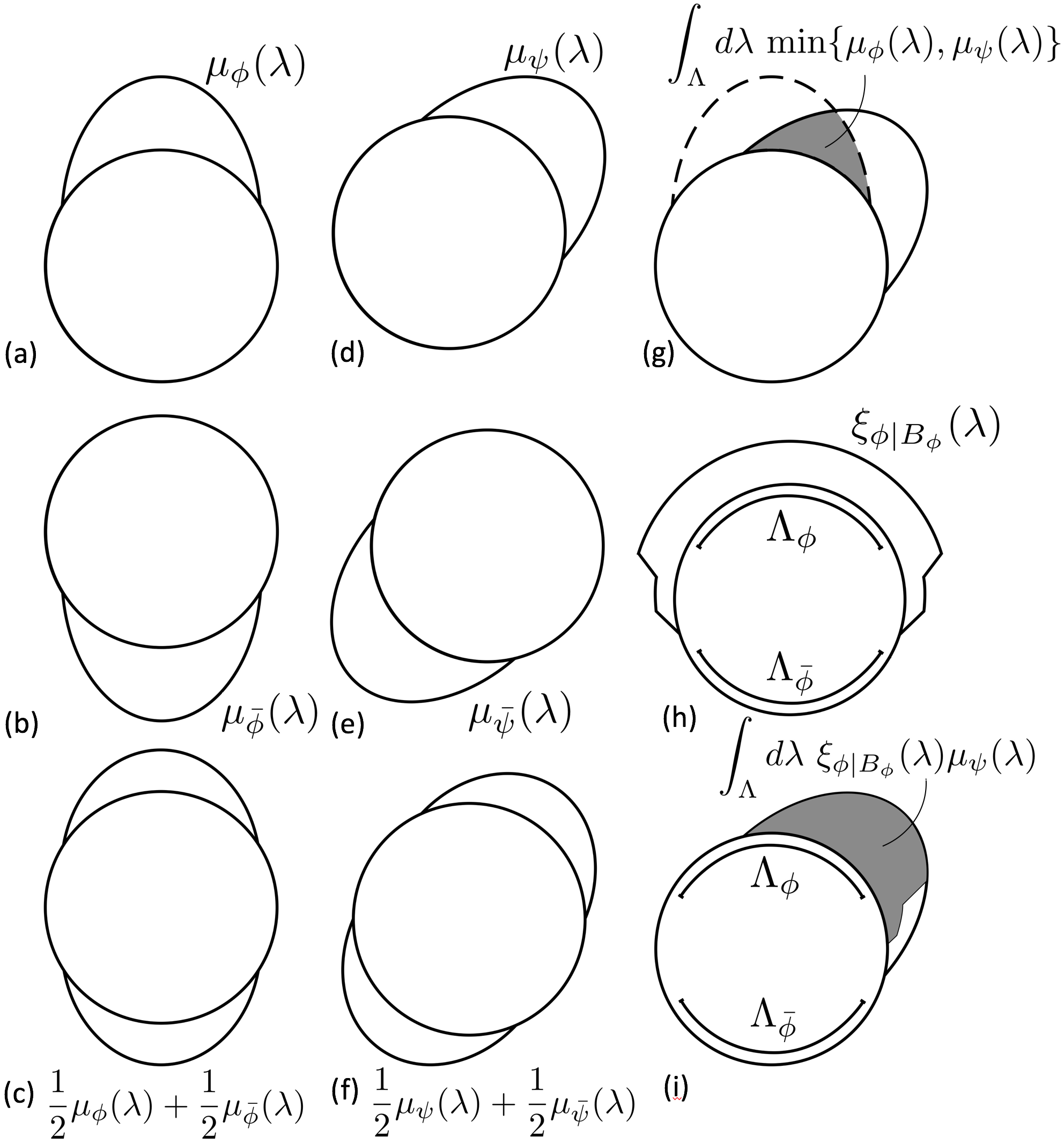}
\caption{
In a contextual model of an MESD scenario: (a)-(f) Epistemic states; (g) Classical overlap between $\mu_{\phi}(\lambda)$ and $\mu_{\psi}(\lambda)$; (h) Response function $\xi_{\phi|B_{\phi}}(\lambda)$, with indication of $\Lambda_{\phi}\equiv \text{supp}[\mu_{\phi}(\lambda)]$ and $\Lambda_{\bar{\phi}}\equiv \text{supp}[\mu_{\bar{\phi}}(\lambda)]$; (i) Confusability defined by $\xi_{\phi|B_{\phi}}(\lambda)$, also with indication of $\Lambda_{\phi}$ and $\Lambda_{\bar{\phi}}$.
}
\label{overlaps1}
\end{figure}

\begin{figure}[!htb] 
\centering
\includegraphics[width=0.5\textwidth]{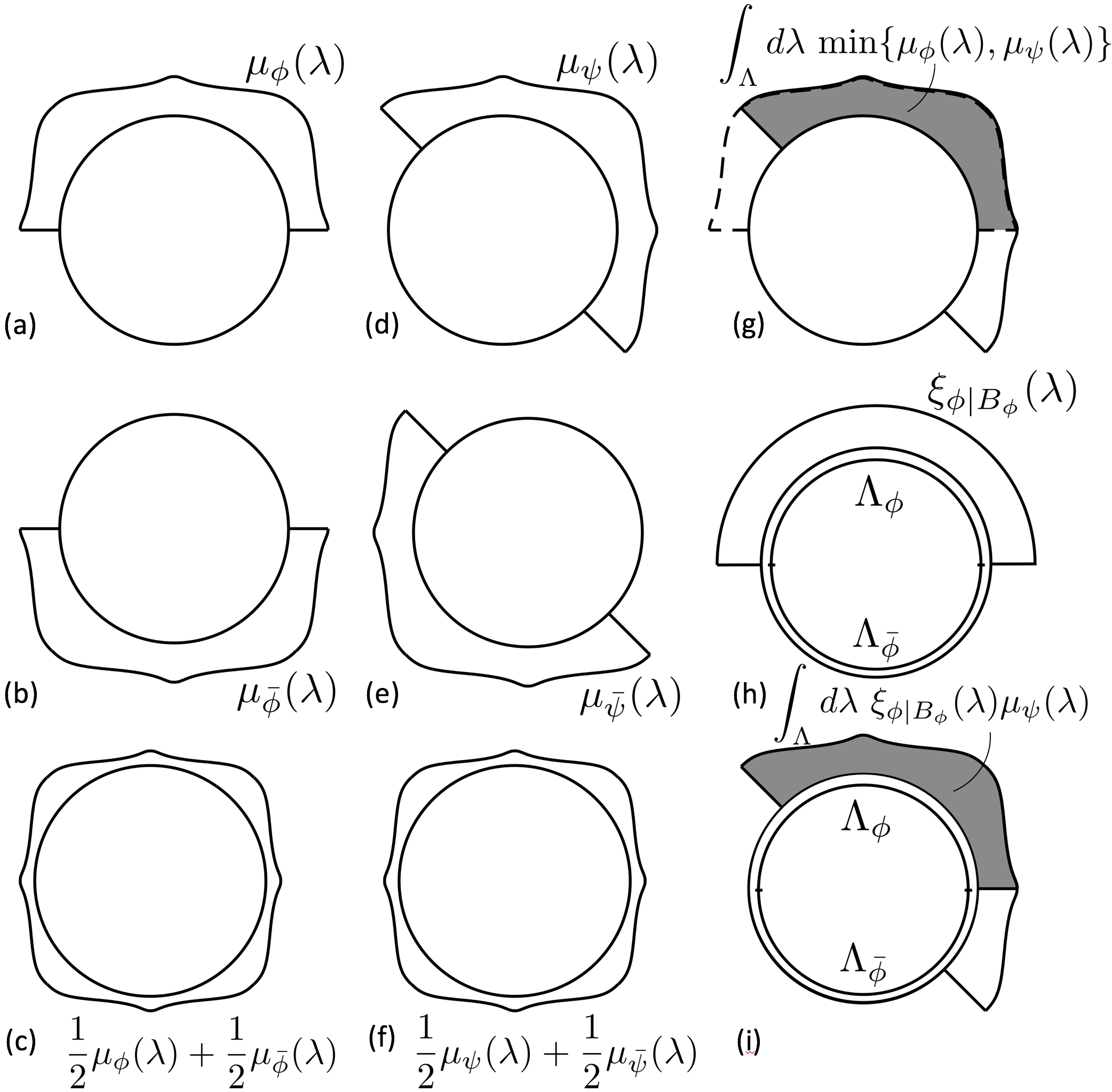}
\caption{
In a noncontextual model of an MESD scenario: (a)-(f) Epistemic states; (g) Classical overlap between $\mu_{\phi}(\lambda)$ and $\mu_{\psi}(\lambda)$; (h) Response function $\xi_{\phi|B_{\phi}}(\lambda)$, with indication of $\Lambda_{\phi}\equiv \text{supp}[\mu_{\phi}(\lambda)]$ and $\Lambda_{\bar{\phi}}\equiv \text{supp}[\mu_{\bar{\phi}}(\lambda)]$; (i) Confusability defined by $\xi_{\phi|B_{\phi}}(\lambda)$, also with indication of $\Lambda_{\phi}$ and $\Lambda_{\bar{\phi}}$..
}
\label{overlaps2}
\end{figure}

The intuitive proof is best summarized graphically, by contrasting a preparation-contextual ontological model, Fig.~\ref{overlaps1}, with a preparation noncontextual ontological model, Fig.~\ref{overlaps2}. For visual simplicity, we have chosen a continuous, 1-dimensional, 
bounded 
ontic state space. We arrange the state space into a circle, so that each point on the circle is a unique ontic state, and epistemic states are represented as probability distributions on the surface of the circle (where the probability density corresponds to the radial height). In each figure, we show the epistemic states for the four preparations and for the two mixed preparations, the classical overlap for two epistemic states, a representative response function, and the confusability generated by that response function. We then show that in the contextual model, the classical overlap and confusability can differ, while in the noncontextual model, they must be identical.

In the ontological model of an MESD scenario shown in Fig.~\ref{overlaps1}, the distributions $\frac{1}{2}\mu_{\phi}(\lambda) + \frac{1}{2}\mu_{\bar{\phi}}(\lambda)$ and $\frac{1}{2}\mu_{\psi}(\lambda) + \frac{1}{2} \mu_{\bar{\psi}}(\lambda)$ are not identical; hence, this model is preparation-contextual. The classical overlap $\int_{\Lambda} d\lambda \  \text{min} \{ \mu_{\phi}(\lambda), \mu_{\psi}(\lambda) \}$ is equal to the area of the shaded region in (g). The response function $\xi_{\phi|B_{\phi}}(\lambda)$ must have value 0 on the support of $\mu_{\bar{\phi}}(\lambda)$ and value 1 on the support of $\mu_{\phi}(\lambda)$, as pictured in (h); however, in the region outside both of these supports, its value is arbitrary, as indicated schematically. Given the response function pictured, the confusability $c_q = \int_{\Lambda} d\lambda \xi_{\phi|B_{\phi}}(\lambda) \mu_{\psi}(\lambda)$ equals the area of the shaded region in (i). One can clearly see that the classical overlap and the confusability need not be the same in a preparation-contextual model.

In the ontological model of an MESD scenario shown in Fig.~\ref{overlaps2}, the distributions $\frac{1}{2}\mu_{\phi}(\lambda) + \frac{1}{2}\mu_{\bar{\phi}}(\lambda)$ and $\frac{1}{2}\mu_{\psi}(\lambda) + \frac{1}{2} \mu_{\bar{\psi}}(\lambda)$ \textit{are} identical; hence, this model is preparation-noncontextual. Furthermore, these two distributions are equal to the unique distribution $\mu_{\mathbb{1}/2}(\lambda)$ (whose support must span the entire ontic state space), and the epistemic states $\mu_{\phi}(\lambda)$, $\mu_{\bar{\phi}}(\lambda)$, $\mu_{\psi}(\lambda)$, and $\mu_{\bar{\psi}}(\lambda)$ must both be equal on their support to $2\mu_{\mathbb{1}/2}(\lambda)$. 
Thus, in a preparation-noncontextual model, the classical overlap is given simply by the integral of $2\mu_{\mathbb{1}/2}(\lambda)$ in the region of common support, as shown by the shaded region in (g). 
Furthermore, preparation noncontextuality implies that the response function $\xi_{\phi|B_{\phi}}(\lambda)$ is 1 on the support of $\mu_{\phi}(\lambda)$ and 0 on \textit{all} other ontic states, as shown in (h). Given this form for the response function, the confusability $c_q = \int_{\Lambda} d\lambda \xi_{\phi|B_{\phi}}(\lambda) \mu_{\psi}(\lambda)$ is given by the area of the shaded region in (i). Clearly, the classical overlap and the confusability are identical in a preparation-noncontextual model.

\subsection{Relation to previous work} \label{relation}

Leifer and Maroney~\cite{Leifer} consider  the assumption that Eq.~\eqref{expforcq} should hold for {\em every} possible pair of quantum states $\phi$ and $\psi$ as a constraint on ontological models that is worthy of investigation in its own right. They term ontological models that satisfy this assumption {\em  maximally $\psi$-epistemic}.  As we noted in Sec.~\ref{intuitive} (and as demonstrated in their article), this assumption follows from preparation noncontextuality (and hence from universal noncontextuality). However, Leifer and Maroney investigate the consequences of making the assumption of maximal $\psi$-epistemicity without also assuming other consequences of universal noncontextuality, in particular, without assuming other consequences of preparation noncontextuality.
 
 
 They establish their no-go theorem for maximal $\psi$-epistemicity (and hence for universal noncontextuality) by demonstrating that maximal $\psi$-epistemicity implies the Kochen-Specker notion of noncontextuality (which is measurement noncontextuality together with the assumption of outcome determinism for sharp measurements), and then relying on the fact that quantum theory does not admit of a Kochen-Specker noncontextual model (the Kochen-Specker theorem).  

Both our article and theirs explore senses in which a pair of quantum states may be said to be ``indistinguishable'', and to what extent some operational counterpart of this indistinguishability can be explained in an ontological model satisfying certain properties.  But there are key differences. As we've noted, the property of ontological models that we focus on is different: we consider the assumption of universal noncontextuality rather than just maximal $\psi$-epistemicity.\footnote{Whereas we believe that the assumption of universal noncontextuality is well motivated (namely, by Leibniz's principle of the identity of indiscernibles), it is unclear to us whether any motivation can be given for maximal $\psi$-epistemicity that is not simultaneously a motivation for universal noncontextuality.  Therefore, unlike Ref.~\cite{Maroney}, we remain unconvinced that the assumption that Eq.~\eqref{expforcq} holds for every pair of quantum states is interesting in its own right.}  
The more important difference between our work and that of Leifer and Maroney, however, is in how we operationalize the notion of indistinguishability.




To explain the difference,
it is useful to
highlight two distinct facts about a pair of nonorthogonal pure quantum states (i.e., a pair $|\psi\rangle$ and $|\phi\rangle$ for which $|\langle \psi | \phi \rangle|^2  >0$): (i) they are not perfectly {\em discriminable}, which is to say that there is no quantum measurement that achieves zero error in the discrimination task, formalized as $s_q>0$, and (ii) they are {\em confusable}, which is to say that the ideal quantum measurement that tests for being in the state $|\phi\rangle$ has a nonzero probability of being passed by the state $|\psi\rangle$, and similarly for $|\phi\rangle$ and $|\psi\rangle$ interchanged, formalized as $c_q >0$.

The determination of the maximum probability of discrimination for a given confusability, that is, the optimal tradeoff relation that holds between $s_q$ and $c_q$, is one of the central results in the field of quantum state estimation. Our work seeks to determine constraints on this tradeoff relation from assumptions about the ontological model. 

Leifer and Maroney, by contrast, do not consider this tradeoff relation, nor the expression for the discriminability of quantum states. Rather, they address (and answer in the negative) the question of whether the degree of confusability of nonorthogonal pure quantum states can be given a particular expression in the ontological model, namely, that of Eq.~\eqref{expforcq}, which asserts that the test associated to the state $|\phi\rangle$ is a test for whether the ontic state $\lambda$ is inside the ontic support of the distribution representing $|\phi\rangle$.\footnote{This is the sort of explanation one obtains in the toy theory model of the single qubit stabilizer subtheory of quantum theory~\cite{toy} or the Kochen-Specker model of a single qubit~\cite{KS}.  Note that this is not the only way to explain the degree of confusability; the response function for $|\phi\rangle$ might be nontrivial outside the ontic support of the distribution representing $|\phi\rangle$ and even indeterministic in that region, and if so, one can have a nonzero confusability even though $\mu_{\phi}$ and $\mu_{\psi}$ have disjoint ontic supports.}
While the expression for the confusability of two quantum states is a feature of their indistinguishability, it is not one that has previously been of interest in the field of quantum state estimation.

Thus, whereas Leifer and Maroney show the impossibility of a particular ontological expression for the confusability from a known no-go result for Kochen-Specker noncontextuality (the Kochen-Specker theorem), we begin with the native phenomenology of minimum-error state discrimination (the quantum tradeoff between $s_q$ and $c_q$), and we derive a novel no-go result for universal noncontextuality from it. 

The form of the tradeoff relation between discriminability and confusability has relevance for quantum information processing tasks that make use of minimum error state discrimination.  For instance, it is used in Ref.~\cite{RudolphSpekkensBC} to derive the tradeoff relation between concealment and bindingness in quantum bit commitment protocols~\cite{bitcommit,LoChauBC}, and such protocols can be used as subroutines in protocols for other tasks, such as strong coin flipping~\cite{AmbainisCF,RudolphSpekkensBC}.  It has also used in the analysis of quantum protocols for the task of oblivious transfer~\cite{Sikora}. 
Our results may be useful, therefore, in determining whether or not the failure of universal noncontextuality is a resource for such tasks.

 Note that because MESD for two pure quantum states is a phenomenon occuring in a two-dimensional Hilbert space (the subspace spanned by the two states) while the Kochen-Specker theorem can only be proven in Hilbert spaces of dimension three or greater, there is no possibility of leveraging facts about Kochen-Specker-uncolourable sets to infer anything about which aspects of MESD resist explanation within a universally noncontextual model.\footnote{The reason there is no possibility of proving the Kochen-Specker theorem with projective measurements in dimension 2 is that no projector appears in more than a single context~\cite{Harrigan,KS}. By contrast, it is known that there are proofs of the failure of preparation noncontextuality that hold even in 2-dimensional Hilbert spaces~\cite{gencontext}, and the proof we have presented here is of this type.}

A final crucial advantage of our approach over that of Ref.~\cite{Leifer} is that it can be used to derive noncontextuality inequalities  that are noise-robust and hence experimentally testable, as we will show in the next section.  Noise-robustness is critical if one hopes to leverage contextuality as a resource in real (hence noisy) implementations of information-processing protocols.

\section{Dealing with noise}

It is important to recognize that the inequality of Eq.~\eqref{eq} is not experimentally testable. To clarify this point, we first draw a distinction between \textit{noncontextuality no-go results} and \textit{noncontextuality inequalities}. A \textit{noncontextuality no-go result} is a proof that no noncontextual model can reproduce certain predictions of quantum theory; as such, a no-go result can contain idealizations (such as perfect correlations) which are justified by quantum theory but which never hold in real experiments. In some cases (as above), a no-go result may derive an inequality on the way to deriving a logical contradiction, but such an inequality may not qualify as a proper noncontextuality inequality. In our usage, a \textit{noncontextuality inequality} makes no reference to the quantum formalism and must not invoke idealized assumptions in its derivation. We give such an inequality for MESD in Section~\ref{sec:ineq}.

The distinction between no-go results and robust inequalities has historical precedent. In his 1964 paper~\cite{Bell}, in deriving an inequality that could be shown to be violated by quantum correlations, Bell assumed an experiment wherein certain pairs of measurements had perfectly correlated outcomes. Such perfect correlations hold for ideal quantum states and measurements, but are never observed in nature. Hence, Bell's 1964 result is a no-go result, with consequences for the interpretation of quantum theory, but the inequality he derives en route to this contradiction does not provide a means of experimentally testing the principle of local causality. In 1969, Clauser, Horne, Shimony, and Holte~\cite{CHSH} derived an inequality without assuming these idealizations. Because their inequality makes no reference to perfect correlations or to any other feature of quantum theory, its violation rules out all locally causal ontological models, independently of the validity of quantum theory. Only inequalities of this type are termed ``Bell inequalities'' in modern usage (so that the inequality in Bell's 1964 paper is not a ``Bell inequality'').

Similarly, Eq.~\eqref{eq} is not a proper noncontextuality inequality because it relies upon the idealization of perfect correlations between which of the states $\ket{\phi}$ or $\ket{\bar{\phi}}$ was prepared and which of the outcomes will occur in the measurement of the $B_{\phi}$ basis (and similarly for $\psi$ and $\bar{\psi}$). 
To get a noncontextuality inequality, we must allow these correlations to be imperfect.  Thus, in Table~\ref{QMtable}, the entries that take the values 0 and 1 must instead be presumed to take the values $\epsilon$ and $1-\epsilon$ respectively, such that  $\epsilon$ becomes a parameter in our noncontextuality inequality which quantifies the degree of imperfection of the correlations.
We then show that quantum mechanics still allows higher success rates for a given confusability than any noncontextual model, even when $\epsilon \neq 0$. 

Before proving this, we first rephrase the scenario as a totally operational prepare-and-measure experiment, with no reference to the quantum formalism (despite the suggestive notation below). This is a necessary first step for deriving any proper noncontextuality inequality. 

\subsection{Operationalizing MESD} \label{operational}

We imagine an experiment involving four preparations $\{ P_{\phi}, P_{\psi},  P_{\bar{\phi}}, P_{\bar{\psi}} \}$ and three binary-outcome measurements, $\{ M_{\phi}, M_{\psi}, M_d\}$, with outcome sets denoted $\{ \phi, \bar{\phi}\}$, $\{ \psi, \bar{\psi}\}$, and $\{ g_{\phi}, g_{\psi} \}$, respectively. An arbitrary data table for such an experiment would contain 12 independent parameters, specifying the probability of the first outcome of each measurement when acting on each preparation (the probability of obtaining the second outcome being fixed by normalization).

However, we wish to study the scenario in which preparations $P_{\phi}$, $P_{\psi}$, $P_{\bar{\phi}}$, and $P_{\bar{\psi}}$ satisfy the following relation: the procedure $P_{\frac{1}{2}\phi + \frac{1}{2}\bar{\phi}}$ defined by sampling from preparations $P_{\phi}$ and $P_{\bar{\phi}}$ uniformly at random (and then forgetting which preparation occurred) is indistinguishable from the similarly defined procedure $P_{\frac{1}{2}\psi + \frac{1}{2}\bar{\psi}}$. We denote this operational equivalence by
\begin{equation} \label{eq:1}
P_{\frac{1}{2}\phi + \frac{1}{2}\bar{\phi}} \simeq P_{\frac{1}{2}\psi + \frac{1}{2}\bar{\psi}}.
\end{equation}
This implies that only 3 of the parameters in each row are independent, so only 9 independent parameters remain.

Previously the operational equivalence of Eq.~\eqref{eq:1} was guaranteed by quantum theory (Eq.~\eqref{mix}), but now we wish to justify it experimentally. In order to do so, one must show that the statistics for $P_{\frac{1}{2}\phi + \frac{1}{2}\bar{\phi}}$ and for $P_{\frac{1}{2}\psi + \frac{1}{2}\bar{\psi}}$ are identical for all measurements. Because the statistics of a tomographically complete set of measurements allows one to predict the statistics for \textit{all} measurements, it suffices to verify this identity for such a tomographically complete set. Accumulating evidence that a given set of measurements is indeed tomographically complete represents the most difficult challenge for an experimental test of noncontextuality (See Refs.~\cite{unwarranted,boot} for a more detailed discussion.).

Note that in a realistic experiment, the four preparations that are realized, called the primary preparations, will not satisfy Eq.~\eqref{eq:1} perfectly. However, this problem can be solved by post-processing these into ``secondary preparations'' which are chosen to enforce this equivalence~\cite{unwarranted,robust}, as discussed in Section~\ref{processing}. 

For this 9-parameter problem, the algorithm we describe in Appendix~\ref{NCnoise} gives the full set of necessary and sufficient noncontextuality inequalities, which we list in Appendix~\ref{fullset}. For now, however, we consider a special case with just three parameters, which captures the essence of minimum error state discrimination. Namely, we assume symmetries that parallel those in the ideal quantum case:
\begin{align} \label{eq:4}
s &\equiv p(g_{\phi} | M_{d},P_{\phi}) =  1-p(g_{\phi} | M_{d},P_{\psi})\\
&= p(g_{\bar{\phi}} | M_{d},P_{\bar{\phi}}) =  1-p(g_{\bar{\phi}} | M_{d},P_{\bar{\psi}}), \nonumber
\end{align}
\begin{align} \label{eq:3}
c &\equiv p(\phi | M_{\phi},P_{\psi}) = p(\psi | M_{\psi},P_{\phi}),\\
 &=p(\bar{\phi} | M_{\bar{\phi}},P_{\bar{\psi}}) = p(\bar{\psi} | M_{\bar{\psi}},P_{\bar{\phi}}) \nonumber
 \end{align}
and
\begin{align} \label{eq:2}
1 - \epsilon &\equiv p(\psi | M_{\psi},P_{\psi}) = p(\phi | M_{\phi},P_{\phi})\\
&= p(\bar{\psi} | M_{\bar{\psi}},P_{\bar{\psi}}) = p(\bar{\phi} | M_{\bar{\phi}},P_{\bar{\phi}}). \nonumber
\end{align}
 We have denoted the three free parameters that remain after imposing the symmetries by $s$, $c$, and $1-\epsilon$, paralleling their ideal quantum counterparts, $s_q$, $c_q$, and $1$, respectively. Just like the operational equivalence, these symmetries will never hold exactly for the primary procedures, but we can enforce them while choosing secondary procedures, as discussed in Section~\ref{processing}. 

The notation $ P_{\phi}, P_{\psi},  P_{\bar{\phi}}$, $P_{\bar{\psi}}$, $M_{\phi}$, $M_{\psi}$, and $M_d$ will henceforth be used to denote the secondary procedures, for which the operational equivalence and symmetries are exact.

The resulting data table, Table~\ref{opertable}, is similar to the ideal scenario of Table~\ref{QMtable}, but contains the noise parameter $\epsilon$ $(1-\epsilon)$ in place of the probability 0 (1). 
\begin{center}
\begin{table} [!htb]
 \begin{tabular}{|c | c | c | c | c|}  
 \hline
  & $P_{\phi}$ &  $P_{\psi}$ & $P_{\bar{\phi}}$ & $P_{\bar{\psi}}$  \\ [0.5ex] 
 \hline 
$\phi|M_{\phi}$ & $1-\epsilon$ & $c$ & $\epsilon$ & $1-c$ \\ [0.2ex] 
 \hline
$\psi|M_{\psi}$ & $c$ & $1-\epsilon$ & $1-c$ & $\epsilon$ \\ [0.2ex] 
 \hline
 $g_{\phi}|M_d$ & $s$ & $1-s$ & $1-s$ & $s$ \\ [0.2ex] 
 \hline
\end{tabular}
\caption{Data table for our operational scenario.} \label{opertable}
\end{table}
\end{center}
Note that for each row, the average of the entries in the $P_{\phi}$ and $ P_{\bar{\phi}}$ columns is $\frac{1}{2}$ (and similarly for $P_{\psi}$ and $ P_{\bar{\psi}}$). Here, this follows from the assumed symmetries, not from the operational equivalence (which specifies that the average of the entries for $P_{\phi}$ and $ P_{\bar{\phi}}$ is the same as the average of the entries for $P_{\psi}$ and $ P_{\bar{\psi}}$, but not necessarily $\frac{1}{2}$); in Table~\ref{QMtable}, the same averaging property is implied by the operational equivalence of each of the two mixtures to the maximally mixed quantum state in Eq.~\eqref{mix} (and redundantly implied by these symmetries).

Finally, we assume that the measurements and outcomes are labeled in the natural way; e.g., the outcome of $M_{\phi}$ that is more likely to occur given the preparation $P_{\phi}$ is $\phi$ rather than $\bar{\phi}$, etc. Then, the data satisfies the constraint that 
\beq \label{labelling}
\epsilon \leq c \leq 1-\epsilon.
\eeq

\section{Noncontextuality inequalities for MESD} \label{sec:ineq}

The operational equivalence relation of Eq.~\eqref{eq:1} together with the assumption of preparation noncontextuality implies via Eq.~\eqref{PNCdefn} that 
\begin{equation} 
\frac{1}{2}\mu_{P_{\phi}}(\lambda) + \frac{1}{2}\mu_{P_{\bar{\phi}}}(\lambda) = \frac{1}{2}\mu_{P_{\psi}}(\lambda) + \frac{1}{2} \mu_{P_{\bar{\psi}}}(\lambda),
\end{equation}
where we have again used the fact that convex mixtures of preparations are represented in an ontological model by the corresponding mixture of epistemic states.
The fact that the ontological model must reproduce Table~\ref{opertable} implies constraints analogous to Eqs.~\eqref{c1}-\eqref{c3} and their kin.  

As we prove in Appendix~\ref{NCnoise}, the tradeoff between $s$, $c$, and $\epsilon$ in any noncontextual model of our operational scenario must satisfy
\beq \label{result}
s \leq 1 - \frac{c - \epsilon}{2}.
\eeq 
In Appendix~\ref{a:QMmodel}, we show that quantum theory allows a tradeoff of
\beq \label{noisyHel}
s = \frac{1}{2} (1 + \sqrt{1-\epsilon + 2\sqrt{\epsilon(1-\epsilon)c(c-1)}+c(2\epsilon-1)}).
\eeq
Thus quantum theory predicts a higher state discrimination success rate for any given $c$ and $\epsilon$ than a noncontextual model allows. One easily verifies that Eq.~\eqref{result} reduces to Eq.~\eqref{eq} in the limit of $\epsilon \rightarrow 0$, and that Eq.~\eqref{noisyHel} reduces to Eq.~\eqref{sdef} in the same limit. It is an open question whether Eq.~\eqref{noisyHel} is the optimal tradeoff that quantum theory allows. We conjecture that it is optimal for pairs of states in a 2-dimensional Hilbert space.

\begin{figure}[h]
\centering
\includegraphics[width=0.4\textwidth]{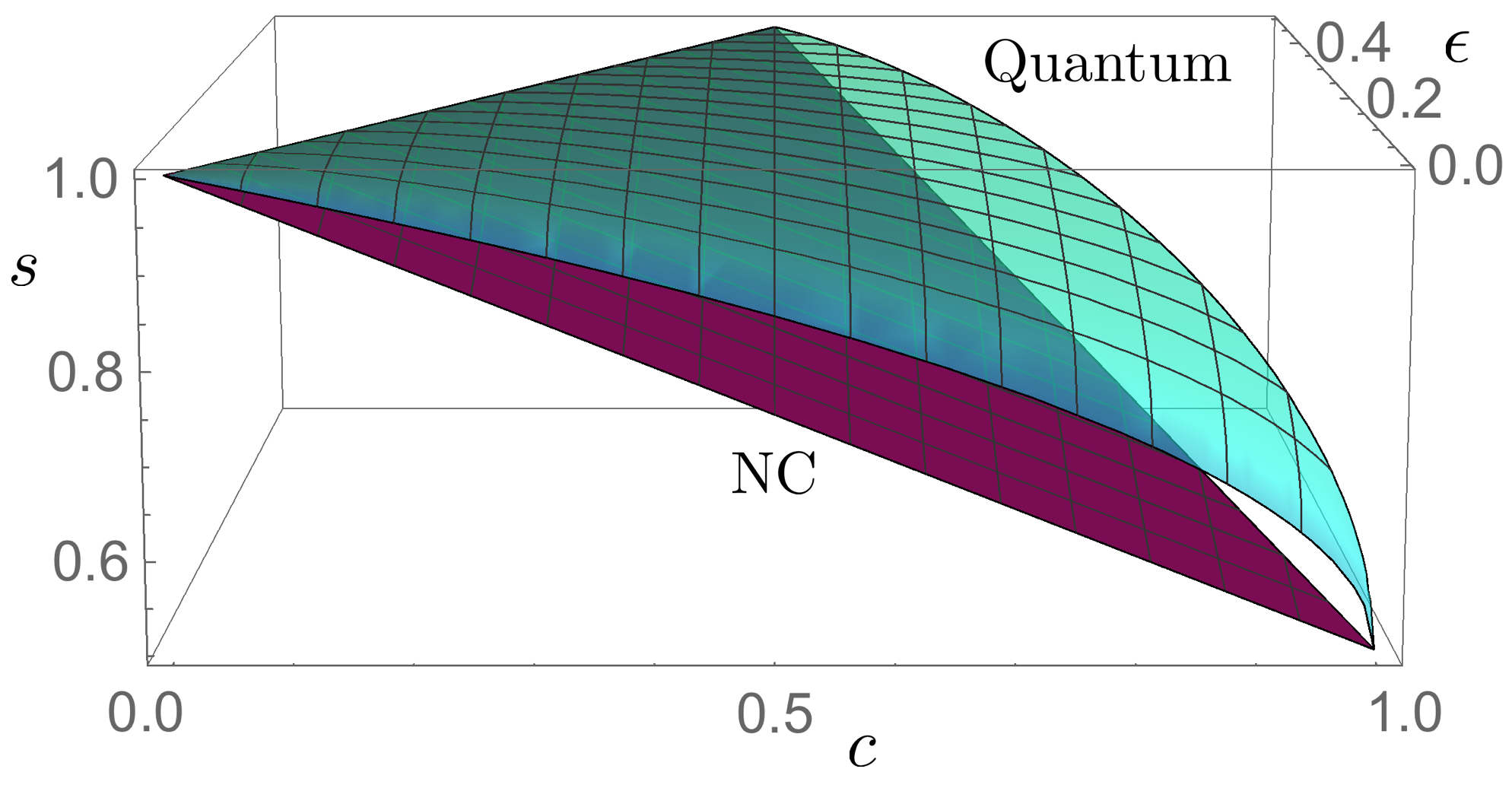}
\caption{Maximum success rate achievable in a noncontextual model (purple surface), and quantumly-acheivable success rate (light blue surface).} \label{QMandNC2}
\end{figure}

The noncontextual and quantum tradeoffs are shown in Fig.~\ref{QMandNC2}. The purple surface represents the triples $(s,c,\epsilon)$ saturating the inequality of Eq.~\eqref{result}, while the light blue surface represents the triples $(s,c,\epsilon)$ corresponding to the quantum success rate of Eq.~\eqref{noisyHel}.

If an experiment generates data having the form of Table~\ref{opertable} and satisfying Eq.~\eqref{eq:1}, and it is found to lie above the purple shaded surface, then one has experimental evidence for the failure of noncontextuality. This evidence is independent of the validity of quantum theory, and signals a contextual advantage for state discrimination, even when one's preparations and measurements are imperfect.

\subsection{Understanding the quantum and noncontextual bounds} \label{understanding}

For both quantum and noncontextual models, we adopt the natural labeling convention described above Eq.~\eqref{labelling}, so that all operational data necessarily satisfies $\epsilon \leq c \leq 1-\epsilon$.  In the $c-\epsilon$ plane of Fig.~\ref{QMandNC2}, these constraints describe a triangular wedge which points into the page. 

In the plane with $\epsilon = 0$, Section~\ref{sec:noiseless} provides an intuitive explanation for the tradeoff relation.

In the plane with $\epsilon = c$, we can see that for both quantum and noncontextual models, the preparations can be perfectly distinguishable, $s=1$. This follows from the fact that the value of $\epsilon$ quantifies the noise in $M_{\phi}$ and $M_{\psi}$, and when $c$ is no larger than $\epsilon$ we can attribute \textit{all} of the confusability to this noise. Explicitly, one can construct a quantum model where preparation $P_{\phi}$ is represented by $\ket{0}\bra{0}$ and $P_{\psi}$ is represented by $\ket{1}\bra{1}$ and where effect $E_{\phi|M_{\phi}}$ is represented by 
$(1-\epsilon)\ket{0}\bra{0} + \epsilon\ket{1}\bra{1}$
 and $E_{\psi|M_{\psi}}$ is represented by $\epsilon\ket{0}\bra{0} + (1- \epsilon)\ket{1}\bra{1}$, which implies that $c=\epsilon$, while $s=1$ for the Helstrom measurement $\{\ket{0}\bra{0}, \ket{1}\bra{1}\}$. Furthermore, since these states and effects are all diagonal in the same basis, we can take the eigenvalues of these to define the conditional probabilities of a noncontextual model which achieves $c=\epsilon$ and $s=1$.


Whenever $c > \epsilon$, however, the noise in $M_{\phi}$ and $M_{\psi}$ cannot explain all of the confusability, and therefore some of this confusability must be explained by the lack of perfect distinguishability of the preparations; that is, in a quantum model, the preparations must be represented by  nonorthogonal states, while in a noncontextual model, they must be represented by overlapping probability distributions.
Thus, the maximum value of $s$ falls away from $1$  as we move away from the $\epsilon = c$ plane. In a noncontextual model, it falls off linearly, interpolating between its value for $\epsilon = c$ and its value for $\epsilon = 0$. The quantum bound falls off more slowly.

\subsection{Robustness to depolarizing noise} \label{sec:depol}

\begin{figure}[!htb] 
\centering
\includegraphics[width=0.5\textwidth]{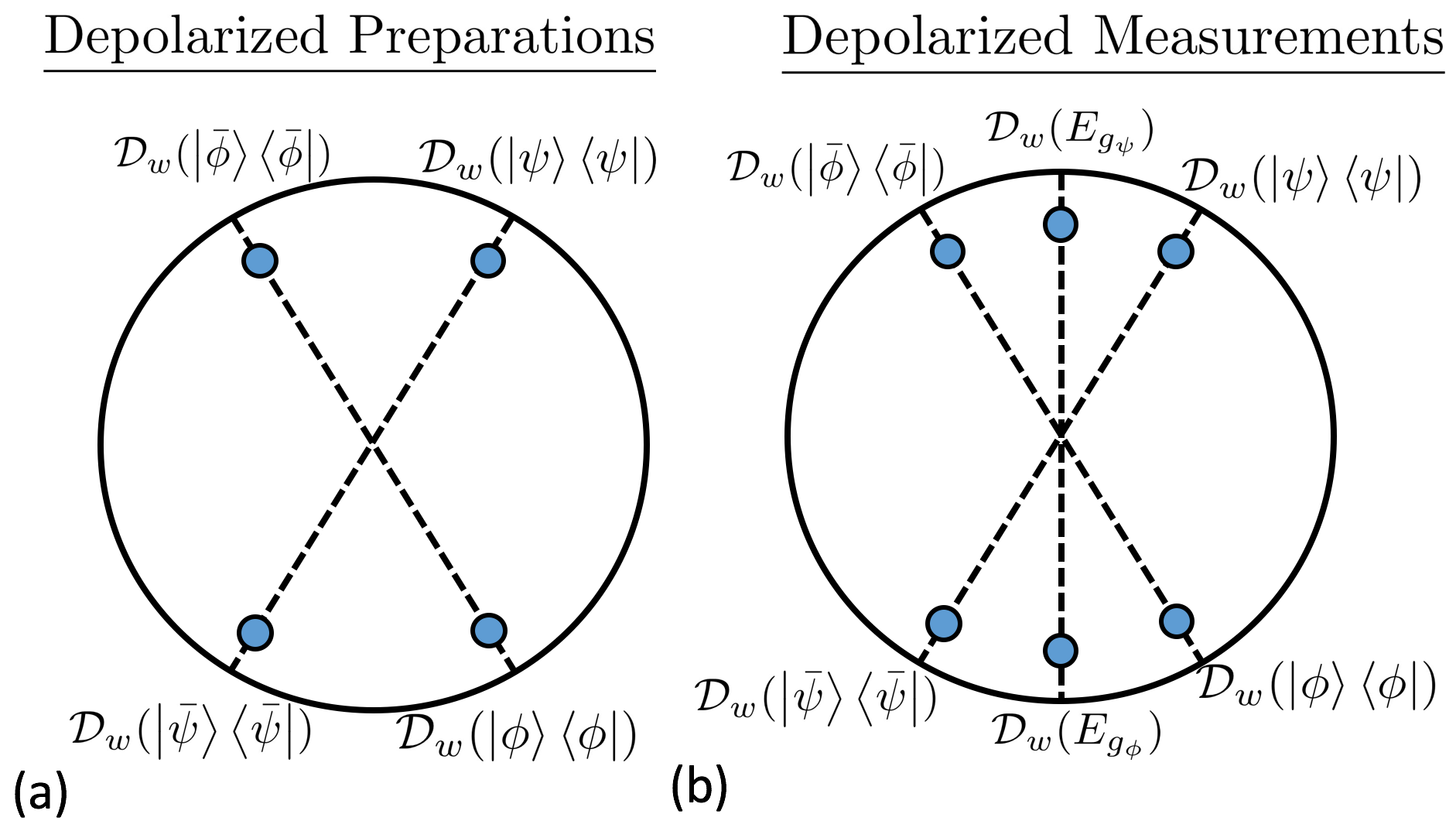}
\caption{The images of the ideal quantum states and effects under a depolarization map for some fixed value of $v$.}
\label{fig:depol}
\end{figure}

We can get a sense for the robustness of our noncontextuality inequalities by considering a specific noise model in quantum theory. Imagine that one's attempts to implement the ideal quantum preparations and measurements are thwarted by a depolarizing channel which has the same noise parameter $v$ for all states and effects:
\begin{align}
\mathcal{D}_v(\rho) = (1-v) \rho + v \frac{\mathbb{1}}{2} \\
\mathcal{D}_v(E_k) = (1-v) E_k + v \frac{\mathbb{1}}{2}.
\end{align}
The resulting states and effects are shown in Fig.~\ref{fig:depol} for some fixed $v$. One can graphically see that this uniform depolarization map generates a new set of states and measurements which satisfy the symmetries and operational equivalence we require. However, if the noise is too large, our noncontextuality inequality will not be violated, as we now show.

This noisy model generates a data table of the form of Table~\ref{opertable} with 
\beq \label{ss}
s 
= \frac{1}{2}+ (1-v)^2((\frac{1}{2}(1+\sqrt{1-c_q}))-\frac{1}{2}),
\eeq
\beq
c = \frac{1}{2}+ (1-v)^2(c_q-\frac{1}{2}),
\eeq
\beq \label{ee}
\epsilon = \frac{1}{2}(1-(1-v)^2).
\eeq
As always, $c_q = |\braket{\phi|\psi}|^2$. 

\begin{figure}[h] 
\centering
\includegraphics[width=0.45\textwidth]{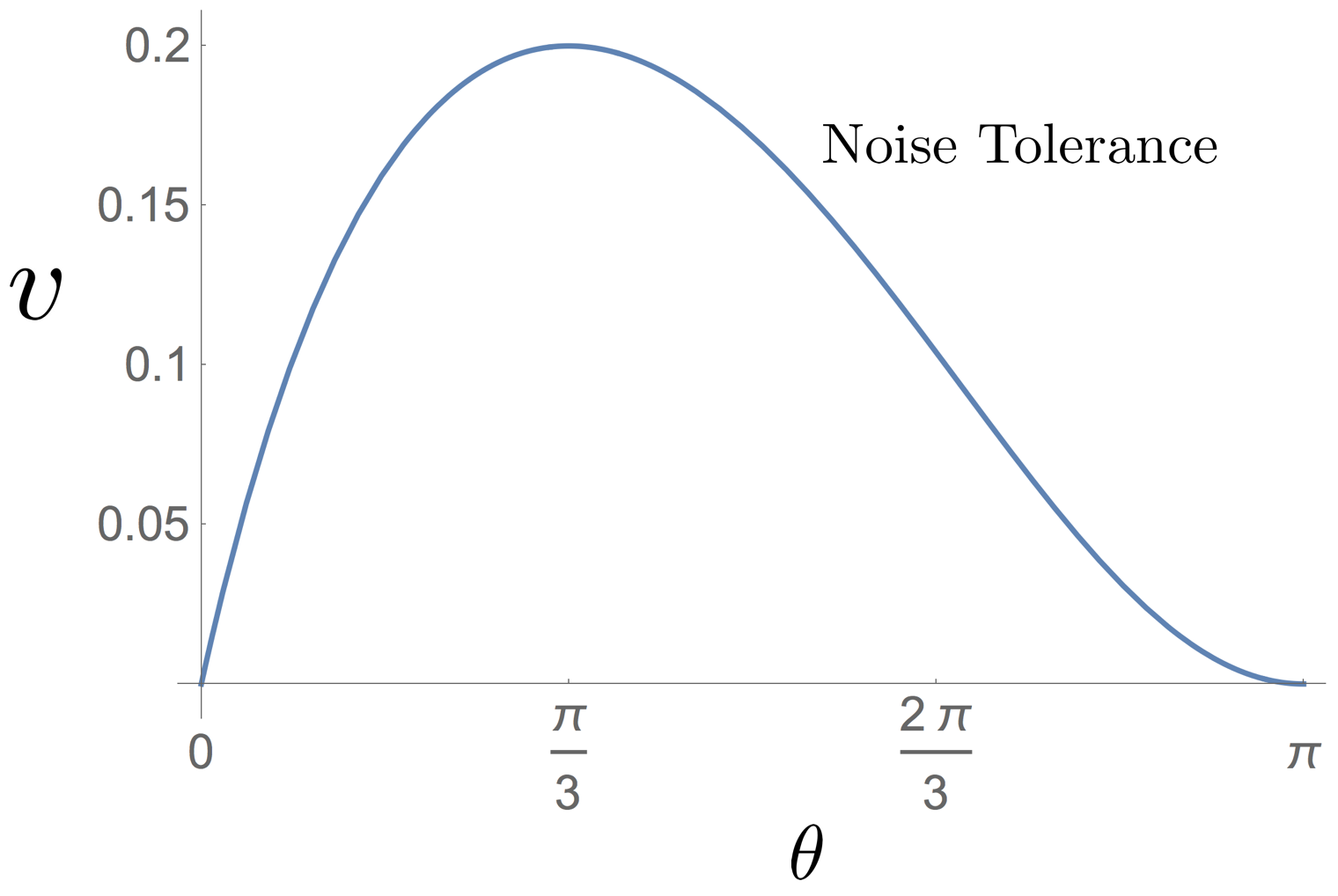}
\caption{The maximum value of the parameter $v$ for the depolarizing noise model that allows a violation of our noncontextuality inequality, as a function of the Bloch sphere angle $\theta$ between the two states.}
\label{minw}
\end{figure}

The maximum level of noise $v$ that still violates our noncontextuality inequality, Eq.~\eqref{result}, is easily calculated as a function of the Bloch sphere angle $\theta$ between the two states (defined by $\text{cos}^2(\frac{\theta}{2})=|\braket{\phi|\psi}|^2$), by substituting Eqs.~\eqref{ss}-\eqref{ee} into Eq.~\eqref{result}:
\beq \label{maxw}
v = 1- \frac{1}{c_q+\sqrt{1-c_q}} = 1-\frac{1}{\text{cos}^2(\frac{\theta}{2})+\text{sin}(\frac{\theta}{2})}.
\eeq
Eq.~\eqref{maxw} is plotted in Fig.~\ref{minw}. For $\theta=0$ or $\theta=\pi$, the noncontextual bound equals the ideal quantum bound, and hence no experiment can violate our noncontextuality inequality at these extremal angles. For all other $\theta$, an experiment with depolarizing noise such that $v \leq 1- \frac{1}{\text{cos}^2(\frac{\theta}{2})+\text{sin}(\frac{\theta}{2})}$ can violate the inequality. The maximum tolerance to noise ($v=0.2$) occurs when $\theta = \frac{\pi}{3}$.

\section{Enforcing symmetries and operational equivalences} \label{processing}


In Section~\ref{operational}, we predicated our noncontextuality inequalities on the exact operational equivalence of Eq.~\eqref{eq:1} and exact operational symmetries of Eq.~\eqref{eq:4}-\eqref{eq:2}, yet we claimed that these idealizations \textit{can} in fact be realized in realistic, noisy experiments. Of course, no experimental data will \textit{directly} satisfy either of these requirements; rather, one performs a post-processing of the data, as originally outlined in~\cite{unwarranted}.   

For pedagogical clarity, we will discuss this data processing under the assumption that the operational theory is quantum theory.   Note, however, that our comments can easily be generalized to the framework of generalized probabilistic theories (defined in Refs.~\cite{Hardy,Barrett}), as demonstrated in Refs.~\cite{unwarranted} and \cite{boot}. Indeed, the analysis {\em must} be performed in this framework if one hopes to directly test the hypothesis of noncontextuality against one's experimental data (i.e., without assuming the validity of quantum theory).

For any set $\frak{P}$ of noisy preparations that has been performed experimentally, one can simulate perfectly the statistics of all other preparations in the convex hull of $\frak{P}$, viewed as points in the quantum state space (here, simply a plane of the Bloch sphere). Similarly, for any set $\frak{E}$ of noisy measurement effects, one can perfectly simulate the statistics of all other effects in the convex hull of $\frak{E}$, viewed as points in the space of valid quantum effects. In~\cite{unwarranted}, this fact was leveraged to simulate exact operational equivalences for a set of ``secondary preparations'' from data on a set of ``primary preparations'' that failed to satisfy the operational equivalences exactly. Here, we leverage this trick to simulate preparations and measurements which simultaneously satisfy our operational equivalence \textit{as well as} the symmetries. We now argue that this can always be done, although if the primary preparations or measurements are too noisy, the resulting simulated data will not violate our inequalities. 

\begin{figure}[!htb] 
\centering
\includegraphics[width=0.5\textwidth]{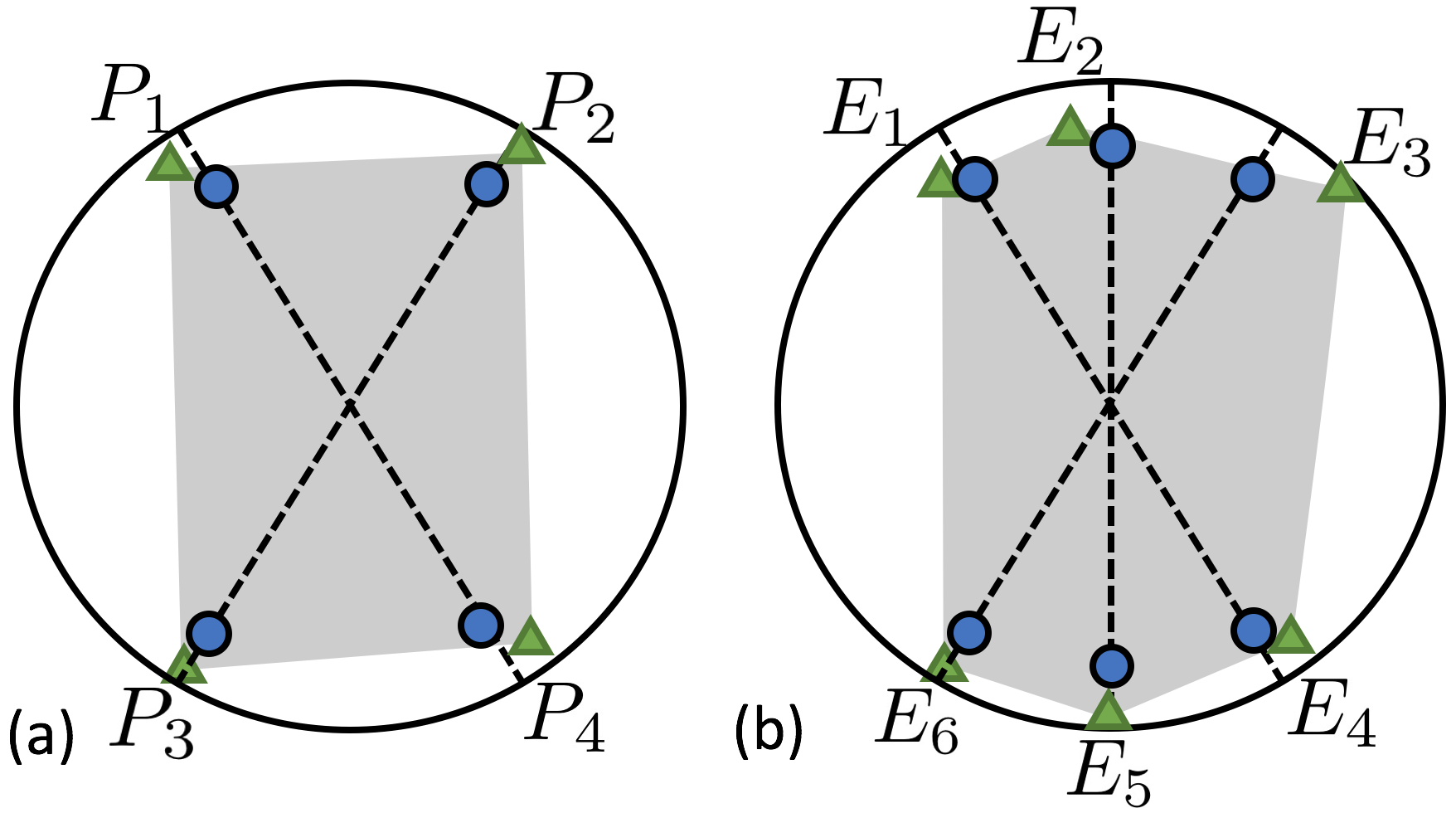}
\caption{(a) If one can perform the four primary preparations $P_1$ to $P_4$ (shown as green triangles), then one can simulate any preparation in their convex hull (shown as a light grey shaded region). In particular, one can simulate secondary procedures that are depolarized versions of the ideal preparations (shown as blue circles like those in Fig.~\ref{fig:depol}). (b) Similarly for the measurements. 
}
\label{secondary}
\end{figure}

As we showed explicitly in Section~\ref{sec:depol}, even a partially depolarized set of states and effects can violate our inequality. Hence, one need only realize experimental sets $\frak{P}$ and $\frak{E}$ which contain in their convex hull the images of our ideal states and effects under the depolarization map $\mathcal{D}_v$ with $v \leq 1- \frac{1}{\text{cos}^2(\frac{\theta}{2})+\text{sin}(\frac{\theta}{2})}$. Then, one can post-process the data obtained from $\frak{P}$ and $\frak{E}$ to obtain a physically meaningful set of data which satisfies the operational equivalence and symmetries that we assumed in the main text, and our inequality will still be violated. Geometrically, this simply means that the primary preparations must have a convex hull which contains the image of the ideal states under a depolarizing map with $v \leq 1-\frac{1}{\text{cos}^2(\frac{\theta}{2})+\text{sin}(\frac{\theta}{2})}$, as pictured in Fig.~\ref{secondary} (and similarly for the measurements, also pictured).

In fact, there are other noisy sets of preparations and measurements besides the depolarized versions of the corresponding ideals which satisfy the operational equivalence and symmetries needed for the noncontextuality inequality to apply. A simple example is states and measurements that are depolarized versions of the ideals that are also rotated in the plane by the same angle. By doing such a rotation, one may be able to simulate a set of states and effects with less depolarization, which then leads to larger violations. In general, there are many sets of states and effects that satisfy our operational equivalence and symmetries. Given a set of primary procedures that one has performed and characterized, finding the states and measurements satisfying our constraints which maximize the violation of our inequality is a straightforward linear program~\cite{unwarranted}.

Leveraging the convex structure of operational theories in order to define secondary laboratory procedures which respect certain theoretical idealizations is a powerful tool which we expect to have broad applicability. To date, this method has been proposed to identify operational procedures which respect exact operational equivalences. What we have just shown is that the method also allows one to enforce natural symmetries which greatly simplify the problem at hand (as evidenced by comparing Eq.~\eqref{result} to the set of inequalities in Appendix~\ref{fullset}). Of course, this tool does not allow one to define laboratory procedures which satisfy \textit{any} desired idealizations; for example, one could never generate a pure state or a sharp measurement effect by convexly mixing the noisy procedures actually performed in the lab. 
We expect future work to continue expanding the range of practical applicability of the technique of secondary procedures.

\section{Isomorphism between MESD and a Bell scenario} \label{Bell2}

Any noncontextuality scenario that makes no assumptions of measurement noncontextuality, and for which there is a \textit{single} mixed preparation whose various ensemble decompositions generate \textit{all} of the operational equivalences of interest, is isomorphic to a related Bell scenario~\cite{parable}. Both of these conditions hold for our MESD scenario, since we do not consider any operational equivalences among the measurements, and the operational equivalences among the preparations are generated by decompositions of a single mixed preparation (e.g. the maximally mixed state in the ideal case). The operational Bell scenario which relates to our MESD scenario is one with two parties, whom we denote by $S$ and $M$ (for reasons that will become apparent), where
$S$ has 2 binary measurements, denoted $S_1$ and $S_2$, and $M$ has 3 binary measurements, denoted $M_1$, $M_2$, and $M_3$. The outcomes (which we denote $s_i$ for $S_i$ and $m_j$ for $M_j$) take values in the set $\{-1,+1\}$.

For such a scenario, the set of constraints which define the local set of correlations is given by positivity inequalities, $p(s_i m_j|S_i M_j) \ge 0$, the normalization condition $\sum_{s_i m_j}p(s_i m_j | S_i M_j) \ge 0$, and the CHSH inequalities~\cite{CHSH} (applied to any of the 3 possible pairings of 2 measurement settings on $S$ with 2 measurement settings on $M$)~\cite{faacets}. As we will show, the bound on our MESD success rate follows under our assumed symmetries from the CHSH inequality 
\beq \label{ineq1}
\left\langle s_1 m_1 \right\rangle + \left\langle s_1 m_3 \right\rangle + \left\langle s_2 m_1 \right\rangle - \left\langle s_2 m_3 \right\rangle \leq 2
\eeq

where
\begin{align} \label{exp}
\left\langle s_i m_j \right\rangle & = \, \sum_{s_i m_j} s_i m_j p(s_i m_j | S_i M_j)\\ 
& = \, 2p(s_i = m_j| S_i M_j) -1. \nonumber
\end{align}

The connection between this Bell scenario and our MESD scenario is most easily seen in the ideal quantum realization.
Imagine that the two parties share a maximally entangled state $\ket{\Phi^+}_{SM} = \frac{1}{\sqrt{2}} (\ket{00}_{SM} + \ket{11}_{SM})$ (with $\ket{0}$ and $\ket{1}$ defined so that $\ket{\phi}$ and $\ket{\psi}$ have real coefficients when written in this basis), and imagine that their measurements correspond to the quantum measurements from the main text, as follows:
\begin{align} \label{correspondence}
S_1 &= \{ \ket{\phi}_S\bra{\phi}, \ket{\bar{\phi}}_S\bra{\bar{\phi}} \} \nonumber \\ \nonumber
S_2 &= \{ \ket{\psi}_S\bra{\psi}, \ket{\bar{\psi}}_S\bra{\bar{\psi}} \}\\ 
M_1 &= \{ \ket{\phi}_M\bra{\phi}, \ket{\bar{\phi}}_M\bra{\bar{\phi}} \}\\ \nonumber
M_2 &= \{ \ket{\psi}_M\bra{\psi}, \ket{\bar{\psi}}_M\bra{\bar{\psi}} \}\\ \nonumber
M_3 &=\{ E_M^{g_{\phi}},E_M^{g_{\psi}} \}. \nonumber
\end{align}

We take the $+1$ outcome for each measurement to correspond to the first quantum effect for that measurement. This ideal quantum realization of this Bell scenario is conceptually transformed into our ideal quantum realization of the MESD scenario by viewing a measurement by party $S$ to be a remote preparation (via quantum steering) for party $M$. For example, outcome $+1$ for $S_1$ remotely prepares the state $\ket{\phi}_M$ (which is why we have chosen the notation $S$, for `source'). Similarly, outcome $-1$ for measurement $S_2$ prepares the state $\ket{\bar{\psi}}_M$, and so on. 

Thus, one can verify that in the ideal quantum realization, $s_q$ and $c_q$ become (in our new notation, and assuming the symmetries in Eqs.~\eqref{eq:4}-\eqref{eq:2})
\begin{align} \label{redefine}
&s_q = p(s_1= m_3|S_1 M_3) = 1-p(s_2= m_3|S_2 M_3) \nonumber \\
&c_q = p(s_1= m_2|S_1 M_2) = p(s_2= m_1|S_2 M_1),
\end{align}
while the fact that paired preparations and measurements are perfectly correlated in the ideal quantum realization corresponds to
\beq \label{redefine2}
0 = 1- p(s_1= m_1|S_1 M_1) = 1-p(s_2= m_2|S_2 M_2).
\eeq

Furthermore, the no-signaling condition in the Bell scenario implies the operational equivalence of our MESD scenario. If party $S$ performs measurement $S_1$, the updated state on $M$ will be either $\ket{\phi}$ or $\ket{\bar{\phi}}$ with equal likelihood, and if party $S$ performs measurement $S_2$, the updated state on $M$ will be either $\ket{\psi}$ or $\ket{\bar{\psi}}$ with equal likelihood. In quantum theory, the no-signaling condition implies that the average density operator prepared on $M$ is the same for either choice of measurement by $S$, which is precisely the operational equivalence of Eq.~\eqref{mix}.  

Using Eq.~\eqref{exp}, we can write Eq.~\eqref{redefine} and Eq.~\eqref{redefine2} in terms of expectation values:
\begin{align}
&s_q = \frac{1}{2}(1+\left\langle s_1 m_3 \right\rangle) = \frac{1}{2}(1-\left\langle s_2 m_3 \right\rangle) \nonumber  \\
&c_q = \frac{1}{2}(1+\left\langle s_1 m_2 \right\rangle) = \frac{1}{2}(1+\left\langle s_2 m_1 \right\rangle)\\ \nonumber 
&0=\frac{1}{2}(1-\left\langle s_1 m_1 \right\rangle) = \frac{1}{2}(1-\left\langle s_2 m_2 \right\rangle).
\end{align}

Rewriting Eq.~\eqref{ineq1} in terms of $s_q$ and $c_q$ instead of expectation values, one obtains 
\beq
s_q \leq 1-\frac{c_q
}{2},
\eeq
recovering Eq.~\eqref{eq}, our bound for the success rate in state discrimination. 

Because both the Bell scenario and our MESD scenario are operationally defined, one can also make the translation without assuming the ideal quantum realizations. In a realistic operational scenario, $\epsilon$ will be nonzero, and one obtains
\begin{align}
&s = p(s_1= m_3|S_1 M_3) = 1-p(s_2= m_3|S_2 M_3) \nonumber  \\
&c = p(s_1= m_2|S_1 M_2) = p(s_2= m_1|S_2 M_1)\\ \nonumber 
&\epsilon = 1- p(s_1= m_1|S_1 M_1) = 1-p(s_2= m_2|S_2 M_2).
\end{align}

Rewriting Eq.~\eqref{ineq1} in terms of $s$, $c$, and $\epsilon$ instead of expectation values, one obtains
\beq
s \leq 1-\frac{c-\epsilon}{2},
\eeq
recovering Eq.~\eqref{result}, our bound for the success rate in state discrimination. 

Due to the redundancies induced by our assumed symmetries, Eq.~\eqref{result} follows also from the CHSH inequality
\beq \label{ineq2}
\left\langle s_1 m_2 \right\rangle + \left\langle s_1 m_3 \right\rangle + \left\langle s_2 m_2 \right\rangle - \left\langle s_2 m_3 \right\rangle \leq 2,
\eeq
by the same logic.
More generally, if we do not assume any symmetries, then there are no redundant inequalities. If we furthermore do not assume the natural labeling constraint (Eq.~\eqref{labelling}), then the full polytope of local correlations for this Bell scenario~\cite{faacets} (and described just above Eq.~\eqref{ineq1}) is isomorphic to the full polytope of noncontextual correlations for our MESD scenario.

\section{Future directions}

We have identified a quantitative feature of minimum-error state discrimination in quantum theory that fails to admit of a noncontextual model. We have derived noncontextuality inequalities that delimit the tradeoff between success rate, error rate, and confusability in state discrimination, independently of the validity of quantum theory. 

Our results show that contextuality is a resource for state discrimination, even in realistic, noisy experiments. This suggests many directions for future research. One important question is how our results translate into advantages for quantum information processing tasks which have MESD as a sub-routine. Because many such tasks (e.g., key distribution) consider consecutive measurements on the system, this research program would require further analysis regarding the consequences of noncontextuality for experiments involving sequential measurements~\cite{PP1,PP2,AWV}. 

It would also be interesting to generalize these results to other types of state discrimination, such as unambiguous state discrimination. Indeed, one can easily derive a relevant no-go theorem. The challenge is to define an operational notion of ``unambiguous'' given that no measurement yields truly unambiguous knowledge in the presence of noise. Once this challenge is met, it should be straightforward to apply the general algorithm we have introduced in this article in order to derive the noncontextuality inequalities for this scenario. Understanding the relation between noncontextuality and other kinds of state discrimination should translate into new kinds of quantum advantages for information processing tasks.

\textit{Acknowledgment}--- D.S. thanks Elie Wolfe and Ravi Kunjwal for useful discussions, Elie Wolfe for his Fourier Motzkin Elimination code, and Marius Krumm and Edwin Tham for useful comments on a draft of the paper. This research was supported by a Discovery grant of the Natural Sciences and Engineering Research Council of Canada and by Perimeter Institute for Theoretical Physics. Research at Perimeter Institute is supported by the Government of Canada through the Department of Innovation, Science and Economic Development Canada and by the Province of Ontario through the Ministry of Research, Innovation and Science.

\appendix

\section{Proof of noncontextuality no-go theorem for MESD} \label{sec:noiseless}

Herein we provide an alternative proof of our no-go theorem, Eq.~\eqref{eq}; that is, of the fact that the inequality
\beq
s_q \leq 1-\frac{c_q}{2}
\eeq
must be satisfied for any $s_q$ and $c_q$ arising in a noncontextual model that reproduces the data in Table~\ref{QMtable} and respects the operational equivalence of Eq.~\eqref{mix}. While the proof provided in the main text uses an intuitive argument that is native to the task of state discrimination, the argument in this section abstracts away from the specific problem at hand, and as such extends naturally to the more general method required for proving Eq.~\eqref{result} (as discussed in Appendix~\ref{NCnoise}).

First, we allow the ontological model to have an ontic state space of arbitrary form, and we allow the response functions to be outcome-indeterministic. Second, we show that for any such model, there exists a simpler ontological model which is equally general, but which has only 8 ontic states and has response functions that are purely outcome-deterministic. Third, we show that two of these ontic states are superfluous if $B_d$ is optimal for state discrimination. Fourth, we show that the forms of the epistemic states are greatly constrained by their perfectly predictable responses on the corresponding measurements. Fifth, we parametrize the set of possible epistemic states as probability distributions over the remaining 6 ontic states in accordance with these constraints. Sixth, we calculate the values of $s_q$ and $c_q$ in terms of these response functions and epistemic states. Finally, we impose preparation noncontextuality and eliminate the unobserved variables to obtain the optimal tradeoff between $s_q$ and $c_q$.

As one ranges over the ontic states in our ontological model, the vector $(\xi_{\phi|B_{\phi}}(\lambda),\xi_{\psi|B_{\psi}}(\lambda),\xi_{g_{\phi}|B_d}(\lambda))$ of valid probability assignments to our three binary basis measurements defines a unit cube. The most obvious ontological model would have one $\lambda$ for each possible probability assignment (including the indeterministic ones), defining an ontic state space isomorphic to the unit cube. The epistemic states in such an outcome-indeterministic model would be arbitrary normalized probability densities over this set of ontic states (that is, all the interior points of the cube).

However, we can always simplify matters without loss of generality by decomposing each non-extremal probability value assignment into extremal assignments. (These extremal points are outcome-deterministic if and only if there are no nontrivial constraints from measurement noncontextuality, but this is indeed the case here.) 

Let us define a variable $\kappa$ which runs over the eight extremal points in the cube of ontic states. Then, there exists a $p(\kappa|\lambda)$ such that $\xi_{k|M}(\lambda) = \sum_{\kappa} \xi_{k|M}(\kappa) p(\kappa|\lambda)$. We can thus write any observable probability $p(k | M,P)$ as 
\begin{equation}
p(k | M,P) = \int_{\Lambda} \xi_{k|M}(\lambda) \mu_P(\lambda) d\lambda =  \sum_{\kappa} \xi_{k|M}(\kappa) \mu_P(\kappa)
\end{equation}
where $\mu_P(\kappa) \equiv \int_{\Lambda} d\lambda \  p(\kappa|\lambda) \mu_P(\lambda)$. This construction lets us write observed probabilities in terms of extremal value assignments by effectively moving uncertainty into the new state distributions $\mu_P(\kappa)$. 

We sometimes simplify the notation by letting the distributions and response functions be vectors of probabilities indexed by the ontic states $\kappa$; e.g.
\beq
p(k | M,P) =  \sum_{\kappa} \xi_{k|M}(\kappa) \mu_P(\kappa) = \vec{\xi}_{k|M} \cdot \vec{\mu}_P .
\eeq

We thus convert an outcome-indeterministic model over a continuum of ontic states (the unit cube) to an outcome-deterministic model over just 8 ontic states (its vertices), without any loss of generality. The vertices $\kappa_1$ to $\kappa_8$ correspond to the deterministic triples
\begin{align}
\begin{split}
\Big(\xi_{\phi|B_{\phi}}(\kappa)&,\xi_{\psi|B_{\psi}}(\kappa),\xi_{g_{\phi}|B_d}(\kappa)\Big) \in \\
 \{ &(0,0,0),(0,0,1),(0,1,0),...,(1,1,1) \},
\end{split}
\end{align}
so the three response functions are
\beq \label{resp2}
\vec{\xi}_{\phi|B_{\phi}}  =    \begin{bmatrix}
0\\0\\0\\0\\1\\1\\1\\1
\end{bmatrix}\!, \ \
\vec{\xi}_{\psi|B_{\psi}}  =    \begin{bmatrix}
0\\0\\1\\1\\0\\0\\1\\1
\end{bmatrix}\!, \ \
\vec{\xi}_{g_{\phi}|B_d}  =    \begin{bmatrix}
0\\1\\0\\1\\0\\1\\0\\1
\end{bmatrix}.
\eeq

In fact, if we assume $B_d$ is optimal, the fourth and fifth of these value assignments will never occur. Consider for example the triple $(1,0,0)$ (which occurs for $\kappa_5$). Since $\xi_{\phi|B_{\phi}}(\kappa_5) = 1$, the state cannot have been $\bar{\phi}$. Since $\xi_{\psi|B_{\psi}}(\kappa_5) = 0$, the state cannot have been $\psi$. Thus, we know the state must have been $\phi$ or $\bar{\psi}$; in either case, the winning strategy is for $B_d$ to return the outcome $g_{\phi}$. Therefore the winning strategy has $ \xi_{g_{\phi}|B_d}(\kappa_5) = 1$, and thus the triple $(1,0,0)$ never occurs in the winning strategy. Similar logic applies to the triple $(0,1,1)$, and hence we need not consider these two assignments \footnote{These assumptions for $B_d$ ensure that the relationship we derive between $s_q$ and $c_q$ will saturate the bound on $s_q$ implied by any noncontextual model. If we had not used this argument, we would obtain the same relationship, but only as a bound on $s_q$, not as the saturating equality. This is easily verified explicitly, e.g. by taking $\epsilon = 0$ in Appendix~\ref{NCnoise} below. However, including two more ontic states requires considerably more algebra.}. The remaining value assignments are 
\begin{align}
\begin{split}
& \Big(\xi_{\phi|B_{\phi}}(\kappa),\xi_{\psi|B_{\psi}}(\kappa),\xi_{g_{\phi}|B_d}(\kappa)\Big) \in \\ 
\{ &(0,0,0),(0,0,1),(0,1,0),(1,0,1),(1,1,0),(1,1,1) \}.
\end{split}
\end{align}

Thus six ontic states are sufficient for describing our experiment: one for each remaining deterministic assignment. It follows that the vectors representing each of the three response functions are:
\beq \label{resp1}
\vec{\xi}_{\phi|B_{\phi}}  =    \begin{bmatrix}
0\\0\\0\\1\\1\\1
\end{bmatrix}\!, \ \
\vec{\xi}_{\psi|B_{\psi}}  =    \begin{bmatrix}
0\\0\\1\\0\\1\\1
\end{bmatrix}\!, \ \
\vec{\xi}_{g_{\phi}|B_d}  =    \begin{bmatrix}
0\\1\\0\\1\\0\\1
\end{bmatrix}.
\eeq

We can constrain the most general form of the epistemic states using the perfect predictability of measurements $B_{\phi}$ and $B_{\psi}$ on their corresponding states. Namely, recalling Eq.~\eqref{ODSM} and the form of the response functions, $\xi_{\phi|B_{\phi}}(\kappa)$, $\xi_{\psi|B_{\psi}}(\kappa)$, $\xi_{\bar{\psi}|B_{\psi}}(\kappa) \equiv 1-\xi_{\psi|B_{\psi}}(\kappa)$, and $\xi_{\bar{\phi}|B_{\phi}}(\kappa) \equiv 1-\xi_{\phi|B_{\phi}}(\kappa)$, we can see that our epistemic states must have the form 
\beq \label{states1}
\vec{\mu}_{\phi}   =    \begin{bmatrix}
0\\0\\0\\a_4\\a_5\\a_6
\end{bmatrix}\!, \,  
\vec{\mu}_{\bar{\phi}}   =    \begin{bmatrix}
a_1\\a_2\\a_3\\0\\0\\0
\end{bmatrix}\!, \, 
\vec{\mu}_{\psi}   =    \begin{bmatrix}
0\\0\\b_3\\0\\b_5\\b_6 
\end{bmatrix}\!, \, 
\vec{\mu}_{\bar{\psi}}   =    \begin{bmatrix}
b_1\\b_2\\0\\b_4\\0\\0 
\end{bmatrix},
\eeq
where normalization requires that $a_4+a_5+a_6=1$, and so on.

The definitions of $c_q$ and $s_q$ in Eqs.~\eqref{defc} and~\eqref{defs} translated into our ontological model become
\beq
c_q = \vec{\mu}_{\psi} \cdot \vec{\xi}_{\phi|B_{\phi}} = \vec{\mu}_{\phi} \cdot \vec{\xi}_{\psi|B_{\psi}} = 1- \vec{\mu}_{\bar{\psi}} \cdot \vec{\xi}_{\phi|B_{\phi}} = 1- \vec{\mu}_{\bar{\phi}} \cdot \vec{\xi}_{\psi|B_{\psi}},
\eeq
\beq \label{defn s}
s_q = \vec{\mu}_{\phi} \cdot \vec{\xi}_{g_{\phi}|B_d} = 1- \vec{\mu}_{\psi} \cdot \vec{\xi}_{g_{\phi}|B_d} = 1- \vec{\mu}_{\bar{\phi}} \cdot \vec{\xi}_{g_{\phi}|B_d} =  \vec{\mu}_{\bar{\psi}} \cdot \vec{\xi}_{g_{\phi}|B_d}.
\eeq
Taking these dot products using the vectors in Eq.~\eqref{resp1} and Eq.~\eqref{states1} gives 
\beq
c_q = b_5+b_6 = a_5+a_6 = 1-b_4 = 1-a_3
\eeq
and
\beq
s_q = a_4 + a_6 = 1 - b_6 = 1-a_2 = b_2 + b_4.
\eeq

Because the epistemic states must be normalized, it follows that $b_5+b_6 = 1-b_3$, $ a_5+a_6 = 1-a_4$, $a_4 +a_6= 1-a_5$, and $b_2+b_4 = 1-b_1$. Substituting these four expressions, we obtain
\beq
c_q = 1-b_3 = 1-a_4 = 1-b_4 = 1-a_3
\eeq
and
\beq
s_q = 1-a_5= 1 - b_6 = 1-a_2 = 1-b_1,
\eeq
and hence $b_3 = a_4 = b_4=a_3$ and $a_5=b_6=a_2=b_1$. 

Let us take $s_q=1-a_2$ and $c_q=1-a_3$. If there were no more constraints, then $a_2$ and $a_3$ could range from 0 to 1 independently, and $s_q$ and $c_q$ could take any values from 0 to 1. By imposing preparation noncontextuality, however, we have  
\beq \label{mixedPNC}
\vec{\mu}_{ \frac{\mathbb{1}}{2}} = \frac{1}{2} \begin{bmatrix}
a_1\\a_2\\a_3\\a_4\\a_5\\a_6
\end{bmatrix} =  \frac{1}{2}
\begin{bmatrix}
b_1\\b_2\\b_3\\b_4\\b_5\\b_6 
\end{bmatrix}\!,
\eeq
This implies $b_i = a_i$ for all $i$. Since $a_1+a_2+a_3 =1$ from normalization, $a_1=b_1$ from preparation noncontextuality, and $b_1=a_2$ as derived above, we also have $2a_2+a_3=1$ and hence $c_q=2a_2$. Finally, writing $s_q$ in terms of $c_q$ yields 
\beq \label{idealt}
s_q = 1 - \frac{c_q}{2}.
\eeq

\section{Proof of noncontextuality inequality for MESD} \label{NCnoise}

Herein we prove our noncontextuality inequality, Eq.~\eqref{result}; that is, we prove that
\beq
s \leq 1 - \frac{c - \epsilon}{2}
\eeq
must be satisfied for any $s$, $c$, and $\epsilon$ arising in a noncontextual model that reproduces data in Table~\ref{opertable} and respects Eq.~\eqref{eq:1}.

First, we use the arguments of Appendix~\ref{sec:noiseless} to write down an ontological model with 8 ontic states and purely outcome-deterministic response functions. Second, we parametrize the set of possible epistemic states for this second model in accordance with preparation noncontextuality. Third, we calculate expressions for $s$, $c$, and $\epsilon$ in terms of these response functions and epistemic states. These manipulations reduce the problem to a small set of linear equalities and inequalities over unobserved and observed variables. Finally, we eliminate the unobserved variables to obtain inequalities concerning only the observed variables $s$, $c$, and $\epsilon$.

Exactly as before, we can convert a general, outcome-indeterministic model over a continuum of ontic states (the unit cube) to an outcome-deterministic model over just 8 ontic states (its vertices), without any loss of generality. (As before, this is simply a mathematical construction, and in no way commits us to a fundamental principle of outcome-determinism.) The vertices of the unit cube, $\kappa_1$ to $\kappa_8$, again correspond to the deterministic triples 
\begin{align}
\begin{split}
\Big(\xi_{\phi|M_{\phi}}(\kappa)&,\xi_{\psi|M_{\psi}}(\kappa),\xi_{g_{\phi}|M_d}(\kappa)\Big) \in \\
 \{ &(0,0,0),(0,0,1),(0,1,0),...,(1,1,1) \},
\end{split}
\end{align}
and the three response functions are again
\beq \label{resp2}
\vec{\xi}_{\phi|M_{\phi}}  =    \begin{bmatrix}
0\\0\\0\\0\\1\\1\\1\\1
\end{bmatrix}\!, \ \
\vec{\xi}_{\psi|M_{\psi}}  =    \begin{bmatrix}
0\\0\\1\\1\\0\\0\\1\\1
\end{bmatrix}\!, \ \
\vec{\xi}_{g_{\phi}|M_d}  =    \begin{bmatrix}
0\\1\\0\\1\\0\\1\\0\\1
\end{bmatrix}.
\eeq
(In a more general situation in which measurement noncontextuality is also leveraged, there will be linear constraints on this set of response functions, and the extremal response functions will no longer all be outcome-deterministic. In this case, one can still explicitly enumerate the finite set of extremal response functions by taking the intersection of the linear constraints with the above cube of value assignments. These extremal points modify the specific form of Eq.~\eqref{resp2}, and our methods would proceed largely unchanged.)

Each preparation generates a probability distribution over $\kappa$, so we can write the epistemic states as
\beq \label{states}
\vec{\mu}_{P_{\phi}}   =    \begin{bmatrix}
a_1\\a_2\\a_3\\a_4\\a_5\\a_6\\a_7\\a_8
\end{bmatrix}\!, \ \  
\vec{\mu}_{P_{\psi}}   =    \begin{bmatrix}
b_1\\b_2\\b_3\\b_4\\b_5\\b_6\\b_7\\b_8, 
\end{bmatrix}\!, \ \ 
\vec{\mu}_{P_{\bar{\phi}}}   =    \begin{bmatrix}
c_1\\c_2\\c_3\\c_4\\c_5\\c_6\\c_7\\c_8, 
\end{bmatrix}\!, \ \
\vec{\mu}_{P_{\bar{\psi}}}   =    \begin{bmatrix}
d_1\\d_2\\d_3\\d_4\\d_5\\d_6\\d_7\\d_8, 
\end{bmatrix}\!,
\eeq
where the parameters in each vector are positive and sum to 1.

Dot products between a vector in Eq.~\eqref{resp2} and a vector in Eq.~\eqref{states} can produce any set of observable statistics, and thus constitute a general ontological model for our measurements and preparations. The values of $(s,c,\epsilon$) that we can observe in a noncontextual model with our assumed symmetries, however, are restricted by the above constraints, all of which we repeat here for convenience.

Eqs.~\eqref{prob1} and~\eqref{norm1} imply that for all four preparations,
\begin{equation} \label{prob2}
\forall \kappa: 0 \leq [ \vec{\mu}_P ]_{\kappa} \leq 1
\end{equation}
and
\begin{equation} \label{norm2}
\sum_k [\vec{\mu}_P]_k = 1.
\end{equation}

Eq.~\eqref{eq:1} gives
\beq
\vec{\mu}_{P_{\phi}} + \vec{\mu}_{P_{\bar{\phi}}} = \vec{\mu}_{P_{\psi}} + \vec{\mu}_{P_{\bar{\psi}}}.
\eeq

Eqs.~\eqref{eq:4}-\eqref{eq:2} are, respectively,
\beq \label{defn s}
\vec{\mu}_{P_{\phi}} \cdot \vec{\xi}_{g_{\phi}|M_d} = 1- \vec{\mu}_{P_{\psi}} \cdot \vec{\xi}_{g_{\phi}|M_d}  = s.
\eeq
\beq
\vec{\mu}_{P_{\psi}} \cdot \vec{\xi}_{\phi|M_{\phi}} = \vec{\mu}_{P_{\phi}} \cdot \vec{\xi}_{\psi|M_{\psi}} = c,
\eeq
\beq
\vec{\mu}_{P_{\psi}} \cdot \vec{\xi}_{\psi|M_{\psi}} = \vec{\mu}_{P_{\phi}} \cdot \vec{\xi}_{\phi|M_{\phi}} = 1- \epsilon,
\eeq

Eq.~\eqref{labelling} gives
\beq \label{label}
\epsilon \leq c \leq 1-\epsilon.
\eeq

Eqs.~\eqref{prob2}-\eqref{label} define a set of constraints over the variables $s, c, \epsilon, a_i, b_i, c_i$, and $d_i$ (where $i\in\{1,2,...,8\}$). Although the parameters $a_i, b_i, c_i, d_i$ in our epistemic states are not observable, constraints upon them (Eqs.~\eqref{prob2} and~\eqref{norm2}) have consequences for the set of possible triples $(s,c,\epsilon)$. Finding the set of inequalities over only $(s,c,\epsilon)$ that is implied by the full set of linear equalities and inequalities of Eqs.~\eqref{prob2}-\eqref{label} is algebraically tedious by hand, but straightforward using the well-known Fourier Motzkin Elimination algorithm, which returns our result 
\beq \label{result2}
s \leq 1 - \frac{c - \epsilon}{2}.
\eeq

It is worth noting that the technique for deriving noncontextuality inequalities we have introduced here, insofar as it reduces to a convex hull problem, is an instance of the problem of quantifier elimination. Recent work in quantum foundations has seen increasing use of quantifier elimination algorithms, in noncontextuality~\cite{Anirudh,PeresMermin} as well as other scenarios.  Fourier-Motzkin elimination, which is appropriate for problems wherein the dependence on the variables to be eliminated is linear, has been used to derive Bell inequalities~\cite{Budroni2012}, and also recently, to derive Bell-like inequalities for novel causal scenarios~\cite{ChavesGrossLuft,Chaves,inflation}.  In Ref.~\cite{inflation}, where the problem is reduced to what is known as the classical marginals problem---that of determining whether a given set of distributions on various subsets of a set of variables can arise as the marginals of a single joint distribution over all of the variables---this problem can be solved by performing quantifier elimination on the probabilities in the joint distributions using convex hull algorithms. Nonlinear quantifier elimination using 
 cylindrical algebraic decomposition has also found application in deriving Bell-like inequalities in simple scenarios~\cite{LeeSpekkens,Chaves}. We anticipate that these more general techniques for quantifier elimination will ultimately also find applications to the derivation of noncontextuality inequalities.


\section{Noisy quantum realization which violates our noncontextuality inequality} \label{a:QMmodel}

We now sketch a quantum realization of the MESD scenario for any given values of $c$ and $\epsilon$ satisfying the assumed symmetries and operational equivalences and violating our noncontextuality inequality for all values of $c$ and $\epsilon$. (The ideal quantum realization of the MESD scenario, given earlier, was defined only for $\epsilon=0$.)

There is no general technique for finding the set of all data tables achievable in quantum theory for some prepare-and-measure scenario. For some cases (e.g., Bell tests), this set can be approximated efficiently via the Navascues-Pironio-Acin hierarchy~\cite{NPA}. For situations with multiple preparations or additional constraints, no such method exists yet. 

However, we can apply our understanding from Section~\ref{understanding} to construct a quantum model which recovers Eq.~\eqref{noisyHel}, which we conjecture is optimal for qubits. 
Namely, because we want to find the maximum value of $s$ consistent with a given $c$ and $\epsilon$, we should attribute as much of the confusability as possible to noise in the $M_{\phi}$ and $M_{\psi}$ measurements, and only attribute the remainder of the confusability to nonorthogonality of the states. As such, in this section we allow the effects $E_{\phi}$, $E_{\psi}$, $E_{\bar{\phi}}$, and $E_{\bar{\psi}}$ to be noisy POVM elements (unlike in Appendix~\ref{sec:noiseless}, where $E_{\phi}$ denoted a projector onto $\ket{\phi}$, and so on).

\begin{figure}[h] 
\centering
\includegraphics[width=0.4\textwidth]{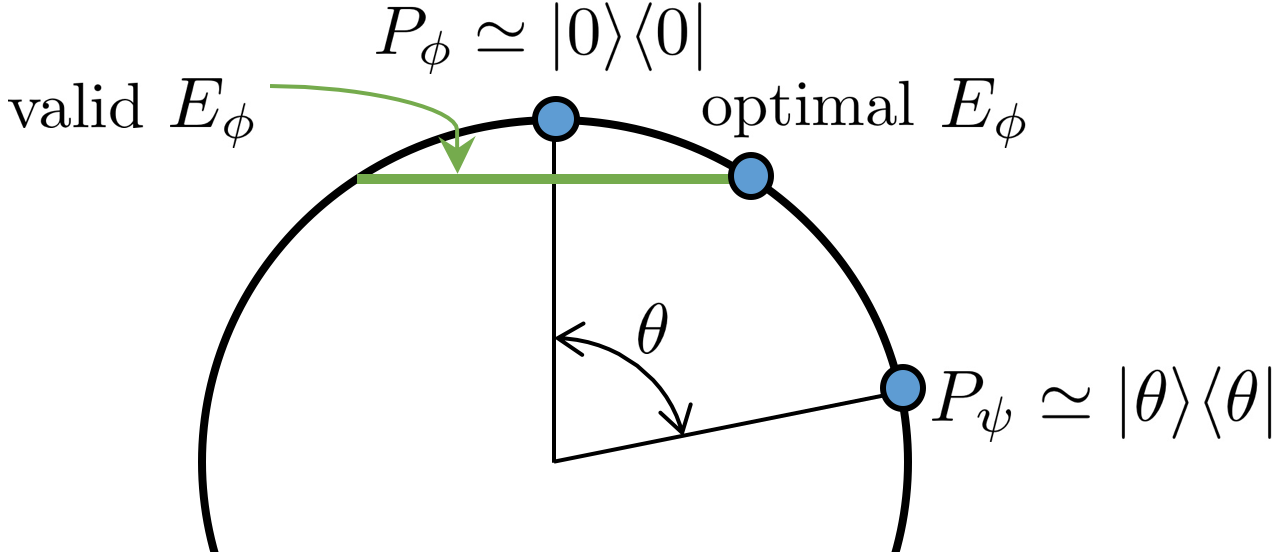}
\caption{Sketch of the quantum model which yields Eq.~\eqref{noisyHel}.}
\label{QMmodel}
\end{figure}

Imagine $P_{\phi}$ prepares state $\ket{0}$ on the Bloch sphere and $P_{\psi}$ prepares a pure state $\ket{\theta}$ rotated by $\theta \in [0,\pi]$ with respect to $\ket{0}$ in the $X-Z$ plane. We will specify the value of $\theta$ later. Within this plane, the effect $E_{\phi}$ must lie on the green line shown in Fig.~\ref{QMmodel}, since only these effects imply $\bra{0} E_{\phi} \ket{0} = 1- \epsilon$. 

The choice of $E_{\phi}$ that yields the maximum confusability is the one on the green line, closest to $\ket{\theta}$ (but not closer to $\ket{\theta}$ than to $\ket{0}$, since that would imply that $c\ge 1-\epsilon$). 
The remainder of the confusability must then be attributed to the nonzero inner product between the two pure states, so $\theta$ is fixed by $\bra{\theta} E_{\phi} \ket{\theta} = c$. Now that the two states are specified, calculating the optimal (Helstrom) probability is a simple quantum calculation whose result gives Eq.~\eqref{noisyHel}, that is
\beq 
s = \frac{1}{2} (1 + \sqrt{1-\epsilon + 2\sqrt{\epsilon(1-\epsilon)c(c-1)}+c(2\epsilon-1)}).
\eeq

The remaining states and effects are completely fixed by the assumed symmetries and operational equivalence.
For a general pair of $c$ and $\epsilon$, this quantum model outperforms the optimal noncontextual model, as seen in Fig.~\ref{QMandNC2}.

\section{Full set of noncontextuality inequalities for MESD without symmetries} \label{fullset}

As promised in Section~\ref{operational}, we now derive the full set of noncontextuality inequalities for our operational MESD scenario when the symmetries of Eqs.~\eqref{eq:4}-\eqref{eq:2} are not assumed. In Table~\ref{opertable2} we show a general data table for 3 binary measurements and 4 preparations which respect our operational equivalence. There are 9 free parameters, since the probabilities in the last column are fixed by those in the first three.

\begin{center}
\begin{table} [!htb]
 \begin{tabular}{|c | c | c | c | c |}  
 \hline
    & $P_{\phi}$ &  $P_{\psi}$ & $P_{\bar{\phi}}$ & $P_{\bar{\psi}}$ \\ [0.5ex] 
 \hline 
$\phi|M_{\phi}$ & $1-\epsilon_{\phi}$ & $c_{\psi}$ & $\epsilon_{\bar{\phi}}$ & $1-\epsilon_{\phi}+\epsilon_{\bar{\phi}}-c_{\psi}$\\ [0.2ex] 
 \hline
$\psi|M_{\psi}$ & $c_{\phi}$ & $1-\epsilon_{\psi}$ & $1-c_{\bar{\phi}}$ & $c_{\phi} -c_{\bar{\phi}} +\epsilon_{\psi}$ \\ [0.2ex] 
 \hline
 $g_{\phi}|M_d$ & $s_{\phi}$ & $1-s_{\psi}$ & $1-s_{\bar{\phi}}$ & $s_{\phi}-s_{\bar{\phi}} -s_{\psi}$\\ [0.2ex] 
 \hline
\end{tabular}
\caption{Data table for our operational scenario with no symmetries assumed. There are 9 free parameters.} \label{opertable2}
\end{table}
\end{center}

The procedure from Appendix~\ref{NCnoise} yields the following set of inequalities over the 9 free parameters, which are necessary and sufficient for the data to have been generated by a noncontextual model respecting operational equivalence Eq.~\eqref{eq:1}:

\begin{align}
&0 \le s_{\phi} \le 1  \\  
&0 \le s_{\bar{\phi}} \le 1 \nonumber \\ 
&0 \le s_{\psi} \le 1 \nonumber \\ 
&0 \le \epsilon_{\phi} \le c_{\phi} \le 1-\epsilon_{\phi} \nonumber \\ 
&0 \le \epsilon_{\bar{\phi}} \le c_{\bar{\phi}} \le 1-\epsilon_{\bar{\phi}} \nonumber \\ 
&0 \le \epsilon_{\psi} \le c_{\psi} \le 1-\epsilon_{\psi} \nonumber \\ 
 &0 \le s_{\phi} - s_{\bar{\phi}} + s_{\psi} \le 1 \nonumber \\ 
&0 \le c_{\phi} - c_{\bar{\phi}} + \epsilon_{\psi} \nonumber \\ 
&0 \le c_{\psi} + s_{\bar{\phi}} - s_{\psi} + \epsilon_{\phi}  \nonumber \\ 
&0 \le c_{\psi} - s_{\bar{\phi}} + s_{\psi} + \epsilon_{\phi}  \nonumber \\ 
&0 \le -c_{\psi} + s_{\phi} +   s_{\psi} + \epsilon_{\bar{\phi}}  \nonumber \\ 
&0 \le c_{\phi} + s_{\bar{\phi}} - s_{\psi} + \epsilon_{\psi}  \nonumber \\ 
&0 \le -c_{\bar{\phi}} + s_{\phi} +   s_{\psi} + \epsilon_{\psi}  \nonumber \\ 
&0 \le c_{\phi} - s_{\bar{\phi}} + s_{\psi} + \epsilon_{\psi}  \nonumber \\ 
&0 \le -c_{\bar{\phi}} +   c_{\psi} + \epsilon_{\phi} + \epsilon_{\psi}  \nonumber \\ 
&0 \le c_{\phi} - c_{\psi} + \epsilon_{\bar{\phi}} + \epsilon_{\psi}  \nonumber \\ 
&0 \le 2 - c_{\psi} - s_{\phi} - s_{\psi} + \epsilon_{\bar{\phi}}  \nonumber \\ 
&0 \le 2 - c_{\bar{\phi}} - s_{\phi} - s_{\psi} + \epsilon_{\psi}  \nonumber \\ 
&0 \le  -c_{\phi} + c_{\bar{\phi}} +   c_{\psi} + \epsilon_{\phi} - \epsilon_{\bar{\phi}} - \epsilon_{\psi}  \nonumber \\ \nonumber
&0 \le 1 - c_{\phi} + c_{\bar{\phi}} -   c_{\psi} - \epsilon_{\phi} + \epsilon_{\bar{\phi}} - \epsilon_{\psi} 
\end{align}

Of course, these inequalities reproduce Eq.~\eqref{result} if the symmetries are now imposed.

In deriving these inequalities, we have assumed the logical labeling of Eq.~\eqref{labelling}. If one drops the labeling condition, then the resulting inequalities are identical to the facets of the Bell polytope discussed in Section~\ref{Bell2} (but have no practical relevance to minimum error state discrimination).

\bibliographystyle{ieeetr}
\bibliography{bib}

\end{document}